\documentclass[journal,twoside]{IEEEtran}

\usepackage{amsmath,amssymb,amsthm}
\usepackage{graphicx,subfigure}
\usepackage{algorithmicx,algorithm,algpseudocode}
\usepackage{cite,bm}
\usepackage[usenames,dvipsnames,svgnames,table]{xcolor}

\usepackage{hyperref}

\newtheorem{theorem}{Theorem}
\newtheorem{lemma}{Lemma}

\newtheorem{proposition}{Proposition}
\theoremstyle{definition}

\newcommand{\cK}{{\cal K}}
\newcommand{\cP}{{\cal P}}
\newcommand{\cD}{{\cal D}}
\newcommand{\cU}{{\cal U}}
\newcommand{\cH}{{\cal H}}
\newcommand{\cQ}{{\cal Q}}

\newcommand{\Tr}{\textrm{Tr}}
\newcommand{\Vect}{\textrm{vec}}
\newcommand{\st}{\textrm{s. t.}}

\renewcommand{\min}{\textrm{min}}
\renewcommand{\max}{\textrm{max}}

\DeclareMathOperator*{\argmax}{arg\,max}
\begin{document}

% ------------------------------------------------------------------------------
% Title.
% ------------------------------------------------------------------------------
\title{Distributed Optimal Beamformers for Cognitive Radios Robust to Channel Uncertainties}

\author{Yu Zhang,~\IEEEmembership{Student~Member,~IEEE,}~Emiliano Dall'Anese,~\IEEEmembership{Member,~IEEE,} \\
and~Georgios B. Giannakis,~\IEEEmembership{Fellow,~IEEE}%
\thanks{Manuscript received January 29, 2012; revised May 31, 2012 and August 15, 2012; accepted August 24, 2012.
This work was supported by QNRF grant NPRP 09-341-2-128. Part of this work was presented at the \emph{37-th International
Conference on Acoustics, Speech, and Signal Processing}, Kyoto, Japan, March 25-30, 2012.}
\thanks{The authors are with the Department of Electrical and Computer Engineering and the Digital Technology Center, University of Minnesota, 200 Union Street SE, Minneapolis, MN 55455, USA. Tel/fax: (612)626-7781/625-2002,~e-mails: {\tt \{yuzhang,emiliano,georgios\}@umn.edu}  %
}
}

% The paper headers
\markboth{IEEE TRANSACTIONS ON SIGNAL PROCESSING (TO APPEAR)}%
{ZHANG \MakeLowercase{\textit{et al.}}: DISTRIBUTED OPTIMAL BEAMFORMERS FOR COGNITIVE RADIOS ROBUST TO CHANNEL UNCERTAINTIES}

\maketitle

% -------------------------------------------------------------------------
% Abstract
% -------------------------------------------------------------------------

\begin{abstract}
Through spatial multiplexing and diversity, multi-input multi-output (MIMO) cognitive radio (CR) networks
can markedly increase transmission rates and reliability, while controlling the interference inflicted to
peer nodes and primary users (PUs) via beamforming. The present paper optimizes the design of transmit- and receive-beamformers
for ad hoc CR networks when CR-to-CR channels are known, but CR-to-PU channels cannot be estimated accurately.
Capitalizing on a norm-bounded channel uncertainty model, the optimal beamforming design is formulated to minimize the overall mean-square error (MSE) from all data streams, while enforcing protection of the PU system when the CR-to-PU channels are uncertain. Even though the resultant optimization problem is non-convex, algorithms with provable convergence to stationary points are developed by resorting to block coordinate ascent iterations, along with suitable convex approximation techniques. Enticingly, the novel schemes also lend themselves naturally to distributed implementations. Numerical tests are reported to corroborate the analytical findings.
\end{abstract}


% -------------------------------------------------------------------------
% Introduction
% -------------------------------------------------------------------------
\section{Introduction}

Cognitive radio (CR) is recognized as a disruptive technology with great potential to enhance
spectrum efficiency. From the envisioned CR-driven applications, particularly promising
is the hierarchical spectrum sharing~\cite{Zhao07}, where CRs opportunistically re-use frequency bands
licensed to primary users (PUs) whenever spectrum vacancies are detected in the time and space dimensions.
Key enablers of a seamless coexistence of CR with PU systems are reliable sensing of the licensed spectrum~\cite{SJEmiGG11,Wang10}, and judicious control of the interference that CRs inflict to PUs~\cite{Zhao07}.
In this paper, attention is focused on the latter aspect.

Recently, underlay multi-input multi-output (MIMO) CR networks have attracted considerable attention thanks to
their ability to mitigate both self- and PU-inflicted interference via beamforming, while leveraging spatial
multiplexing and diversity to considerably increase transmission rates and reliability.
On the other hand, wireless transceiver optimization has been extensively studied in the non-CR
setup under different design criteria~\cite{Cioffi-twc08,Meisam}, and when either perfect
or imperfect channel knowledge is available; see e.g.,~\cite{Ding,Shi}, and references therein.
In general, when network-wide performance criteria such as weighted sum-rate and sum mean-square error (MSE)
are utilized, optimal beamforming is deemed challenging because the resultant
optimization problems are typically non-convex. Thus, solvers assuring even first-order Karush-Kuhn-Tucker (KKT) optimality are appreciated in this context~\cite{Cioffi-twc08, Meisam, Ding}.

In the CR setup, the beamforming design problem is exacerbated by the presence of interference constraints~\cite{Zhao07}.
In fact, while initial efforts in designing beamformers under PU interference constraints were made under the premise
of \emph{perfect} knowledge of the cognitive-to-primary propagation channels~\cite{Rui08,Liangjstsp08,SJK,Scutari},
it has been recognized that obtaining accurate estimates of the CR-to-PU channels is challenging or even impossible.
This is primarily due to the lack of full CR-PU cooperation~\cite{Zhao07}, but also to estimation errors and frequency
offsets between reciprocal channels when CR-to-PU channel estimation is attempted.
It is therefore of paramount importance to take the underlying \emph{channel uncertainties} into account,
and develop prudent beamforming schemes that ensure protection of the licensed users.

Based on CR-to-PU channel statistics, probabilistic interference constraints were employed in~\cite{dallanese-twc11}
for single-antenna CR links. Assuming imperfect knowledge of the CR-to-PU channel, the beamforming design
in a multiuser multi-input single-output (MISO) CR system sharing resources with single-antenna PUs was considered in~\cite{Gharavol-tvt10};
see also~\cite{Zheng} for a downlink setup, where both CR and PU nodes have multiple antennas.
The minimum CR signal-to-interference-plus-noise ratio (SINR) was maximized under a bounded norm constraint capturing uncertainty in the CR-to-PU links.
Using the same uncertainty model, minimization of the overall MSE from all data streams in MIMO ad hoc CR
networks was considered in~\cite{Liang}. However, identical channel estimation errors for different CR-to-PU
links were assumed. This assumption was bypassed in~\cite{KhasibSL11}, where the mutual information was
maximized instead. Finally, a distributed algorithm based on a game-theoretic approach was developed in~\cite{WangSP10}.

The present paper considers an underlay MIMO ad hoc CR network sharing spectrum bands
licensed to PUs, which are possibly equipped with multiple antennas as well.
CR-to-CR channels are assumed known perfectly, but this is not the case for CR-to-PU channels.
Capitalizing on a norm-bounded uncertainty model to capture inaccuracies of the CR-to-PU channel estimates,
a beamforming problem is formulated whereby CRs minimize the overall MSE,
while limiting the interference inflicted to the PUs \emph{robustly}.
The resultant robust beamforming design confronts two major challenges:
\emph{a)} non-convexity of the total MSE cost function;
and, \emph{b)} the semi-infinite attribute of the robust interference constraint,
which makes the optimization problem arduous to manage.  To overcome the second hurdle,
an equivalent re-formulation of the interference constraint as a linear matrix inequality (LMI)
is derived by exploiting the S-Procedure \cite{Boyd}.
On the other hand, to cope with the inherent non-convexity, a cyclic block
coordinate ascent approach~\cite{Bertsekas97} is adopted along with local convex approximation techniques.
This yields an iterative solution of the semi-definite programs (SDPs) involved,
and generates a convergent sequence of objective function values.
Moreover, when the CR-to-CR channel matrices have full column rank, every limit point
generated by the proposed method is guaranteed to be a stationary point of the original non-convex problem.
However, CR links where the transmitter is equipped with a larger number of antennas than the receiver,
or spatially correlated MIMO channels~\cite{Kermoal02}, can lead to beamformers that are not necessarily optimal.
For this reason, a proximal point-based regularization technique~\cite{Rockf76} is also employed to guarantee convergence
to optimal operating points, regardless of the channel rank and antenna configuration.
Similar to~\cite{Ding,Cioffi-twc08,Meisam,Shi,Gharavol-tvt10,Rui08,Liangjstsp08,SJK,Scutari,WangSP10,Zheng,PalomarCL03,Liang,KhasibSL11,Chiang-tsp11,Poor09},
perfect time synchronization is assumed at the symbol level.

Interestingly, the schemes developed are suitable for \emph{distributed} operation,
provided that relevant parameters are exchanged among neighboring CRs. The algorithms can also be implemented in an \emph{on-line} fashion
which allows adaptation to (slow) time-varying propagation channels. In this case, CRs do not necessarily wait for the iterations to converge,
but rather use the beamformer weights as and when they become available. This is in contrast to, e.g.,~\cite{Ding,Cioffi-twc08}
and~\cite{Zheng,Liang,KhasibSL11} in the non-CR and CR cases, respectively, where the relevant problems are solved \emph{centrally} and in a \emph{batch} form.

In the robust beamforming design, the interference power that can be tolerated by
the PUs is initially assumed to be pre-partitioned in per-CR link portions, possibly according to
quality-of-service (QoS) guidelines~\cite{Scutari, WangSP10}. However, extensions of the beamforming design
are also provided when the PU interference limit is \emph{not} divided a priori among CR links.
In this case, primal decomposition techniques~\cite{Bertsekas97} are invoked to dynamically allocate the total interference among CRs.
Compared to~\cite{Liang}, the proposed scheme accounts for different estimation inaccuracies in the CR-to-PU links.

The remainder of the paper is organized as follows.
Section~\ref{sec:systemmodel} introduces the
system model and outlines the proposed robust beamforming design problem.
In Section~\ref{Section:Robust}, the block coordinate ascent solver is developed, along with its proximal point-based alternative.
Aggregate interference constraints are dealt with in Section~\ref{Section:PrimalDecom}, and numerical results
are reported in Section~\ref{sec:numericalresults}.
Finally, concluding remarks are given in Section~\ref{sec:conclusions},
while proofs are deferred to the Appendix.

\emph{Notation}. Boldface lower (upper) case letters represent
vectors (matrices); $\mathbb{H}^{n \times n}, \mathbb{H}^{n \times n}_+,
\mathbb{C}^{n \times n}$ and $\mathbb{R}$ stand for spaces of $n \times n$
Hermitian, $n \times n$ Hermitian positive semidefinite, $n \times n$ complex matrices, and real numbers,
respectively; $(\cdot)^{T}$, $(\cdot)^{*}$, and $(\cdot)^\cH$ indicate transpose,
complex conjugate, and conjugate transpose operations, respectively;
$\Tr\{\cdot\}$ denotes the trace operator, and $\Vect(\mathbf{A})$
the vector formed by stacking the columns of $\mathbf{A}$;
$\|\mathbf{a}\|_{2}$ and $\|\mathbf{A}\|_{F}$ represent the Euclidean norm of $\mathbf{a}$
and the Frobenius norm of $\mathbf{A}$, respectively;
$\mathbf{A}\otimes \mathbf{B}$ is the Kronecker product of $\mathbf{A}$ and $\mathbf{B}$;
$\mathbf{I}_N$ is the $N{\times}N$ identity matrix;
Finally, $\mathbb{E}\{\cdot\}$ denotes the expectation operator,
and $\Re(\cdot)$ stands for the real part of a complex number.

% ----------------------------
% Problem Formulation
% ----------------------------
\section{System Model and Problem Formulation}
\label{sec:systemmodel}
\begin{figure}[t]
\centering
\includegraphics[width=0.4\textwidth]{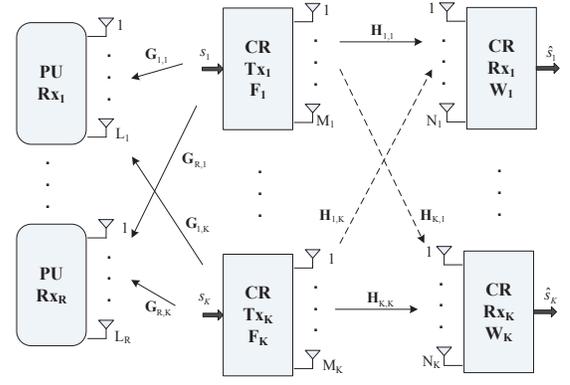}
\caption{The system model for MIMO ad hoc CR networks.}
\label{fig:system}
\end{figure}

Consider a wireless MIMO CR network comprising $K$ transmitter-receiver
pairs $\{U_{k}^t,U_{k}^r\}$. Let $M_k$ and $N_k$, $k \in \cK:=\{1,2,\ldots,K\}$,
denote the number of antennas of the $k$-th transmitter-receiver pair,
as shown in Fig.~\ref{fig:system}.
Further, let $\mathbf{s}_k$ denote the $M_k \times 1$ information
symbol vector transmitted by $U_k^t$ per time slot, with covariance
matrix $\mathbb{E}\{\mathbf{s}_k\mathbf{s}_k^\cH\}=\mathbf{I}_{M_k}$.
In order to mitigate self-interference, transmitter $U_{k}^t$
pre-multiplies $\mathbf{s}_k$ by a transmit-beamforming matrix
$\mathbf{F}_k \in \mathbb{C}^{M_k \times M_k}$; that is, $U_k^t$ actually transmits the
$M_k \times 1$ symbol vector $\mathbf{x}_k:=\mathbf{F}_k\mathbf{s}_k$.
With $\mathbf{H}_{k,j} \in \mathbb{C}^{N_k \times M_j}$ denoting the
$U_{j}^t$ to $U_{k}^r$ channel matrix, the $N_k \times 1$
symbol received at $U_{k}^r$ can be written as
\begin{align}
{\mathbf{y}_{k}} = \mathbf{H}_{k,k} \mathbf{x}_{k}
+ \sum_{j \neq k}\mathbf{H}_{k,j}\mathbf{x}_{j}
+\mathbf{n}_{k}\label{eq:RxSig}
\end{align}
where $\mathbf{n}_{k} \in \mathbb{C}^{N_k}$ is the zero-mean complex
Gaussian distributed receiver noise, which is assumed independent of
$\mathbf{s}_{k}$ and $\{\mathbf{H}_{k,j}\}$, with
covariance matrix $\mathbb{E}\{\mathbf{n}_k \mathbf{n}_k^{\cH}\} =
\sigma _k^2 {\mathbf{I}}_{N_k}$.
%The flat Rayleigh fading is postulated for $\{\mathbf{H}_{k,j}\}$.

Low-complexity receiver processing motivates the use of a linear filter
matrix $\mathbf{W}_k \in \mathbb{C}^{M_k \times N_k}$ at $U_k^r$
to recover $\mathbf{s}_k$ as
\begin{align}
\hat{\mathbf{s}}_{k}:=\mathbf{W}_k\mathbf{y}_k , \quad k \in \mathcal{K}.
\end{align}
Using $\mathbf{W}_k$ at $U_k^r$, the MSE matrix
$\mathbf{E}_{k}:= \mathbb{E}\{\left(\hat{\mathbf{s}}_{k} - \mathbf{s}_{k}\right) \left(\hat{\mathbf{s}}_{k} - \mathbf{s}_{k}\right)^\cH\}$,
which quantifies the reconstruction error, is given by [cf.~\eqref{eq:RxSig}]
\begin{align}
\mathbf{E}_{k} =
\mathbf{W}_{k}\mathbf{A}_{k}\mathbf{W}^{\cH}_{k} -
\mathbf{W}_{k}\mathbf{H}_{k,k}\mathbf{F}_{k} -
\mathbf{F}^{\cH}_{k}\mathbf{H}^{\cH}_{k,k}\mathbf{W}^{\cH}_{k} +
\mathbf{I}_{M_k}
\end{align}
where $\mathbf{A}_{k} := \sum_{j=1}^{K}
\mathbf{H}_{k,j}\mathbf{F}_{j}\mathbf{F}^{\cH}_{j}\mathbf{H}^{\cH}_{k,j}
+ \sigma _k^2\mathbf{I}_{N_k}$. Entry $(i,i)$ of $\mathbf{E}_{k}$
represents the MSE of the $i$-th data stream ($i$-th entry of $\mathbf{s}_k$)
from $U_k^t$ to $U_k^r$, and $\Tr\{\mathbf{E}_{k}\}$ corresponds to the MSE of
$\hat{\mathbf{s}}_{k}$.

To complete the formulation, let $\mathbf{G}_{k} \in \mathbb{C}^{L \times M_k}$ denote
the channel between CR $U_k^t$ and a PU receiver,
possibly equipped with multiple ($L$) antennas\footnote{A
single PU receiver is considered throughout the paper. However,
extension to multiple receiving PUs is straightforward; see also Remark 5.},
and $\iota^{\textrm{max}}$ the maximum instantaneous interference that the PU
can tolerate. As in e.g.,~\cite{Scutari,WangSP10}, suppose that $\iota^{\textrm{max}}$
is pre-partitioned in per-CR
transmitter portions $\{\iota_{k}^{\textrm{max}}\}$,
possibly depending on QoS requirements of individual CR pairs.
Then, the transmit- and receive-beamforming matrices minimizing the
overall MSE can be obtained as (see also~\cite{Ding})
\begin{subequations}
\begin{align}
\textrm{(P1)} \quad
\mathop{\min}\limits_{\{\mathbf{F}_k, \mathbf{W}_k\}^{K}_{k=1}}  &\sum^{K}_{k=1}\Tr\{\mathbf{E}_{k}\}\label{ObjFunc} \\
\st \quad \,  &\Tr\{\mathbf{F}_{k}\mathbf{F}^{\cH}_{k}\} \le p^{\textrm{max}}_{k}, \, k \in \cK~\label{P1powerCsrt}\\
& \Tr\{\mathbf{G}_{k}\mathbf{F}_{k}\mathbf{F}^{\cH}_{k}\mathbf{G}^{\cH}_{k}\} \le \iota_{k}^{\textrm{max}}, \, k \in \cK~\label{P1intfCsrt}
\end{align}
\end{subequations}
where $p_k^{\textrm{max}}$ is the maximum transmit-power of $U_k^t$.

\noindent \textbf{Remark 1} \textit{(Adopted performance metric).}
Among candidate performance metrics, the sum of MSEs
from different data streams is adopted here, which has been widely employed in the beamforming literature; see e.g.,~\cite{Ding,Shi} and references therein. The relationships between MSE, bit error rate (BER) and SINR have been thoroughly considered in~\cite{PalomarCL03}, and further investigated in~\cite{Ding}. Specifically, it has been shown that an improvement in the total MSE naturally translates in a lower BER. Furthermore, the sum of MSEs facilitates derivation of optimal filters, and the equivalence between minimizing the weighted sum of MSEs and maximizing the weighted sum rate has been established in \cite{Cioffi-twc08,Chiang-tsp11}.

Unfortunately, due to lack of explicit cooperation between PU and CR
nodes, CR-to-PU channels $\{\mathbf{G}_{k}\}$ are in general
difficult to estimate accurately. As PU protection must be enforced
though, it is important to take into account the CR-to-PU channel
\emph{uncertainty}, and guarantee that the
interference power experienced by the PU receiver stays below a
prescribed level for \emph{any} possible (random) channel realization~\cite{dallanese-twc11, Zheng}.
Before developing a beamforming approach robust to inaccuracies associated
with channel estimation, problem (P1) is conveniently re-formulated first
in order to reduce the number of variables involved.

% ----------------------------
% Equivalent Problem Formulation
% ----------------------------
\subsection{Equivalent Optimization Problem}

For the sum-MSE cost in~\eqref{ObjFunc}, the optimum $\{\mathbf{W}^{\textrm{opt}}_{k}\}$ will turn out
to be expressible in closed form. To show this, note first that for fixed $\{\mathbf{F}_k\}$, (P1) is convex in $\mathbf{W}_k$, and $\{\mathbf{W}^{\textrm{opt}}_{k}\}$ can be obtained from the first-order optimality conditions. Express the Lagrangian function associated with (P1) as
\begin{align}
\mathcal{L}\left( \cP, \cD \right) = &\sum^{K}_{k=1}\Tr\{\mathbf{E}_{k}\} +
\sum^{K}_{k=1}\lambda_{k} \left(\Tr\{\mathbf{F}_{k}\mathbf{F}^{\cH}_{k}\}-p^{\textrm{max}}_{k}\right) \nonumber\\
& +  \sum^{K}_{k=1} \nu_k
\left(\Tr\{\mathbf{G}_{k}\mathbf{F}_{k}\mathbf{F}^{\cH}_{k}\mathbf{G}^{\cH}_{k}\}-\iota_k^{\textrm{max}}\right)
\end{align}
where $\cP := \{\mathbf{F}_k,\mathbf{W}_k\}_{k=1}^K$ and $\cD := \{\lambda_k,\nu_k\}_{k=1}^K$ collects the primal and dual variables,
respectively. Then, by equating the complex gradient $\partial
\mathcal{L}\left( \cP, \cD \right)/\partial \mathbf{W}_k^{\ast}$ to zero,
matrix $\mathbf{W}_k^{\textrm{opt}}$ is expressed as
\begin{align}
\label{optW}
\mathbf{W}^{\textrm{opt}}_{k} = \mathbf{F}^{\cH}_{k}\mathbf{H}^{\cH}_{k,k}\mathbf{A}_{k}^{-1}, \quad k \in \cK.
\end{align}
Clearly, the optimal set $\{\mathbf{W}^{\textrm{opt}}_{k}\}$ does not depend on channels $\{\mathbf{G}_k\}$, but only on $\{\mathbf{H}_{k,j}\}$.

Substituting $\{\mathbf{W}^{\textrm{opt}}_{k}\}$ into (\ref{ObjFunc}),
and using the covariance $\mathbf{Q}_k := \mathbb{E}\{\mathbf{x}_k\mathbf{x}_k^\cH\} =
\mathbf{F}_{k}\mathbf{F}^{\cH}_{k}$ as a matrix optimization variable, it follows that (P1) can be
equivalently re-written as
\begin{subequations}
\begin{align}
\textrm{(P2)} \quad \mathop{\max}\limits_{\{\mathbf{Q}_{k}\succeq \mathbf{0}\}}
&\sum^{K}_{k=1}u_k\left(\{\mathbf{Q}_{k}\}\right)\label{P3ObjFunc}\\
\st \quad \, &\Tr\{\mathbf{Q}_{k}\} \le p^{\textrm{max}}_{k}, \ k \in \cK~\label{P3TxPower}\\
&\Tr\{\mathbf{G}_{k}\mathbf{Q}_{k}\mathbf{G}^{\cH}_{k}\} \le
\iota_{k}^{\textrm{max}},\ k \in \cK~\label{P3interference}
\end{align}
\end{subequations}
where the per-CR link utility $u_k\left(\{\mathbf{Q}_{k}\}\right)$
is given by
\begin{align}
\label{per-utility}
u_k\left(\{\mathbf{Q}_{k}\}\right) &:= \Tr\left\{\mathbf{H}_{k,k}\mathbf{Q}_{k}\mathbf{H}^{\cH}_{k,k}
\left(\mathbf{H}_{k,k}\mathbf{Q}_{k}\mathbf{H}^{\cH}_{k,k}+\mathbf{R}_{k,k}\right)^{-1}\right\}
\end{align}
with
$\mathbf{R}_{k,k} :=\sum_{i \neq k}\mathbf{H}_{k,i}\mathbf{Q}_{i}\mathbf{H}^{\cH}_{k,i}+\sigma^{2}_{k}\mathbf{I}_{N_k}$.
One remark is in order regarding (P2).

\noindent \textbf{Remark 2} \textit{(Conventional MIMO networks).}
Upon discarding the interference constraints~\eqref{P3interference},
the beamforming problems formulated in this paper along with their
centralized and distributed solvers can be considered also for non-CR
MIMO ad-hoc and cellular networks in downlink or uplink operation.

Channels $\{\mathbf{G}_{k}\}$ must be perfectly known in order to solve (P2).
A robust version of (P2), which accounts for imperfect channel knowledge,
is dealt with in the next section.

\subsection{Robust Interference Constraint}
\label{subSection:RobustInterf}

In typical CR scenarios, CR-to-PU channels are challenging to estimate accurately.
In fact, CR and PU nodes do not generally cooperate~\cite{Zhao07}, thus rendering
channel estimation challenging. To model estimation inaccuracies,
consider expressing the CR-to-PU channel matrix $\mathbf{G}_{k}$ as
\begin{align}
\label{eq:uncertaintymodel}
\mathbf{G}_{k} = \widehat{\mathbf{G}}_{k} + \Delta\mathbf{G}_{k} \, , \quad k \in \cK
\end{align}
where $\widehat{\mathbf{G}}_{k}$ is the estimated channel, which is known at CR transmitter
$U_k^t$, and $\{\Delta\mathbf{G}_{k}\}$ captures the underlying
channel uncertainty~\cite{Zheng, KhasibSL11}. Specifically, the error matrix $\Delta\mathbf{G}_{k}$ is assumed to take values from the bounded set
\begin{align}
\mathcal{G}_{k} := \left\{\Delta\mathbf{G}_{k}|\Tr\{\Delta\mathbf{G}_{k}\Delta\mathbf{G}_{k}^{\cH}\}
\le \epsilon_{k}^{2}\right\}, \, k \in \cK
\label{eq:perturbation}
\end{align}
where $\epsilon_{k} > 0$ specifies the radius of
$\mathcal{G}_{k}$, and thus reflects the degree of uncertainty
associated with $\widehat{\mathbf{G}}_{k}$.
The set in~\eqref{eq:perturbation} can be readily extended to
the general ellipsoidal uncertainty model~\cite[Ch.~4]{Boyd}.
Such an uncertainty model properly resembles the case where a time
division duplex (TDD) strategy is adopted by the PU system,
and CRs have prior knowledge of the PUs' pilot sequence(s). But even without training symbols,
CR-to-PU channel estimates can be formed using the deterministic path loss coefficients,
and the size of the uncertainty region can be deduced from fading channel statistics.
Compared to~\cite{dallanese-twc11}, the norm-bounded uncertainty model leads to \emph{worst-case}
interference constraints that ensure PU protection for any realization of the
uncertain portion of the propagation channels.

Based on~\eqref{eq:perturbation}, a robust interference constraint can be written as
\begin{align}
\Tr\{(\widehat{\mathbf{G}}_{k} + \Delta\mathbf{G}_{k})\mathbf{Q}_{k}(\widehat{\mathbf{G}}_{k} + \Delta\mathbf{G}_{k})^{\cH}\} \le
\iota_{k}^{\textrm{max}}, \nonumber \\
\forall~\Delta\mathbf{G}_{k} \in \mathcal{G}_{k}, \, k \in \cK \label{RobustInterf}
\end{align}
and consequently, a robust counterpart of (P2) can be formulated as follows
\begin{subequations}
\begin{align}
\textrm{(P3)} \quad
\mathop{\max}\limits_{\{\mathbf{Q}_{k}\succeq \mathbf{0}\}} &\sum^{K}_{k=1}u_k\left(\{\mathbf{Q}_{k}\}\right)\label{P4ObjFunc}\\
\st \quad &\Tr\{\mathbf{Q}_{k}\} \le p^{\textrm{max}}_{k}, \ k \in \cK~\label{P4TxPower}\\
& \hspace{-1.5cm} \Tr\{\mathbf{G}_{k}\mathbf{Q}_{k}\mathbf{G}^{\cH}_{k}\} \le
\iota_{k}^{\textrm{max}}, \, \forall~\Delta\mathbf{G}_{k} \in \mathcal{G}_{k}, \,
k \in \cK.~\label{P4Robust}
\end{align}
\end{subequations}

Clearly, once $\{\mathbf{Q}^{\textrm{opt}}_{k}\}$ are found by solving (P3),
the wanted $\{\mathbf{W}^{\textrm{opt}}_{k}\}$ can be readily obtained via~\eqref{optW},
since $\mathbf{A}_{k} := \sum_{j=1}^{K} \mathbf{H}_{k,j}\mathbf{Q}_{j}\mathbf{H}^{\cH}_{k,j}
+ \sigma _k^2\mathbf{I}_{N_k}$. However, $\sum_k u_k\left(\{\mathbf{Q}_{k}\}\right)$ is non-convex
in $\{\mathbf{Q}_{k}\}$, and hence (P3) is hard to solve in
general. Additionally, constraints~\eqref{RobustInterf} are not in a
tractable form, which motivates their transformation.
These issues are addressed in the next section. But first, two remarks are in order.

\noindent \textbf{Remark 3}  \textit{(Uncertain MIMO channels).}
In an underlay hierarchical spectrum access setup, it is very challenging (if not impossible) for the CRs to obtain accurate estimates of the CR-to-PU channels. In fact, since the PUs hold the spectrum license, they have no incentive to feed back CR-to-PU channel estimates to the CR system~\cite{Zhao07}. Hence, in lieu of explicit CR-PU cooperation, CRs have to resort to crude or blind estimates of their channels with PUs. On the other hand, sufficient time for training along with sophisticated estimation algorithms render the CR-to-CR channels easier to estimate. This explains why similar to relevant works~\cite{dallanese-twc11,Gharavol-tvt10,Zheng,Liang,KhasibSL11,WangSP10},  CR-to-CR channels are assumed known, while CR-to-PU channels are taken as uncertain in CR-related optimization methods.
Limited-rate channel state information that can become available e.g., with quantized CR-to-CR channels~\cite{Ding,Shi,dallanese-twc11}, can be considered in future research but goes beyond the scope of the present paper.

\noindent \textbf{Remark 4}  \textit{(Radius of the uncertainty region).}
In practice, radius and shape of the uncertainty region  have to be tailored to the specific channel estimation approach implemented at the CRs, and clearly depend on the second-order channel error statistics. For example, if $\Delta\mathbf{G}_{k}$ has zero mean and covariance matrix $\mathbf{\Sigma}_{\Delta\mathbf{G}_{k}} = \hat{\sigma}^2_{k} \mathbf{I}$, where
$\hat{\sigma}^2_{k}$ depends on the receiver noise power, and the transmit-power of the PU (see, e.g.~\cite{Biguesh06}), then the radius of the uncertainty region can be set to $\epsilon_k^2 = \kappa \xi_k \hat{\sigma}^2_{k}$, where $\xi_k$ denotes the path loss coefficient, and $\kappa > 0$ a parameter that controls how strict the PU protection is. Alternatively, the model $\epsilon_k^2 =  \kappa \hat{\sigma}^2_k\|\hat{\mathbf{G}}_k\|_F^2$ can be utilized~\cite{WangSP10}. If $\mathbf{\Sigma}_{\Delta\mathbf{G}_{k}} \neq \hat{\sigma}^2_{k} \mathbf{I}$, then the uncertainty region can be set to $\mathcal{G}_{k} := \left\{\Delta\mathbf{G}_{k}|\Tr\{\Delta\mathbf{G}_{k} \mathbf{\Sigma}_{\Delta\mathbf{G}_{k}}^{-1} \Delta\mathbf{G}_{k}^{\cH}\}
\le \kappa \right\}$~\cite{Poor09} and the robust constraint~\eqref{RobustInterf} can be modified accordingly. Similar models are also considered in~\cite{Davidson08}.

% ------------------------------------------------------------------------------------
% Distributed Robust CR Beamforming
% ------------------------------------------------------------------------------------
\section{Distributed Robust CR Beamforming}
\label{Section:Robust}

% ----------------------------
%\subsection{Block coordinate ascent with local convex approximation}
% ----------------------------

To cope with the non-convexity of the utility function~\eqref{P4ObjFunc}, a
block-coordinate ascent solver is developed in this section. Define first the sum of
all but the $k$-th utility as
$f_{k}(\mathbf{Q}_{k},\mathbf{Q}_{-k}) :=
\sum_{j \neq k}u_j$, with $\mathbf{Q}_{-k} := \{\mathbf{Q}_{j}|j \neq k\}$.
Notice that $u_k(\cdot)$ is concave and $f_k(\cdot)$ is convex in $\mathbf{Q}_{k}$;
see Appendix~\ref{Appendix:convexfunc} for a proof.
Then, (P3) can be regarded as a difference of convex functions (d.c.) program,
whenever only a single variable $\mathbf{Q}_{k}$ is optimized and $\mathbf{Q}_{-k}$ is kept fixed.
This motivates the so-termed concave-convex procedure~\cite{Yuille03}, which belongs to
the majorization-minimization class of algorithms~\cite{Hunter04}, to solve problem (P3)
through a sequence of convex problems, one per matrix variable $\mathbf{Q}_{k}$.
Specifically, the idea is to linearize the convex function $f_{k}(\cdot)$ around a feasible point
$\tilde{\mathbf{Q}}_{k}$, and thus to (locally) approximate the objective (\ref{P4ObjFunc})
as (see also~\cite{Meisam} and~\cite{SJK})
\begin{align}
&\sum^{K}_{k=1}u_k\left(\{\mathbf{Q}_{k}\}\right)
= u_k\left(\{\mathbf{Q}_{k}\}\right) + f_{k}(\mathbf{Q}_{k},\mathbf{Q}_{-k})\nonumber \\
& \approx u_k\left(\{\mathbf{Q}_{k}\}\right) +
f_{k}(\tilde{\mathbf{Q}}_{k},\mathbf{Q}_{-k})
+\Tr\left\{\mathbf{D}_{k}^{\cH}(\mathbf{Q}_{k}-\tilde{\mathbf{Q}}_{k})\right\}
\label{eq:lla}
\end{align}
where
\begin{align}
\mathbf{D}_{k} :=
\nabla_{\mathbf{Q}_{k}}f_{k}(\tilde{\mathbf{Q}}_{k},\mathbf{Q}_{-k})
:=\left. \frac{\partial f_k}{\partial \mathbf{Q}_{k}^{*}}
\right|_{\mathbf{Q}_{k}=\tilde{\mathbf{Q}}_{k}}.
\end{align}
Therefore, for fixed $\mathbf{Q}_{-k}$, matrix $\mathbf{Q}_{k}$ can be obtained by solving the following sub-problem
\begin{subequations}
\begin{align}
\textrm{(P4)} \quad
\mathop{\max}\limits_{\mathbf{Q}_{k}}\quad
& u_k\left(\mathbf{Q}_{k},\mathbf{Q}_{-k}\right)
+\Tr\left\{\mathbf{D}_{k}^{\cH}\mathbf{Q}_{k}\right\}~\label{P5ObjFunc}\\
\st \quad \, &\mathbf{Q}_{k}\succeq \mathbf{0}\label{P5PSDCsrt}\\
&\Tr\{\mathbf{Q}_{k}\} \le p^{\textrm{max}}_{k}\label{P5powerCsrt}\\
&\Tr\{\mathbf{G}_{k}\mathbf{Q}_{k}\mathbf{G}^{\cH}_{k}\} \le
\iota^{\textrm{max}}_{k},\ \forall~ \Delta\mathbf{G}_{k} \in
\mathcal{G}_{k}~\label{P5RobustInterf}
\end{align}
\end{subequations}
where  (see Appendix~\ref{Appendix:gradD})
\begin{align}
\mathbf{D}_{k}
&:=\left. -\sum_{j \neq k} \mathbf{H}^{\cH}_{j,k}
\mathbf{B}_{j}^{-1}\mathbf{V}_{j}\mathbf{B}_{j}^{-1}
\mathbf{H}_{j,k}\right|_{\mathbf{Q}_{k}=\tilde{\mathbf{Q}}_{k}},~\label{eq:gradD}\\
\mathbf{B}_{j} &:=\sum_{i=1}^{K}\mathbf{H}_{j,i}\mathbf{Q}_{i}\mathbf{H}^{\cH}_{j,i}+\sigma^{2}_{j}\mathbf{I}_{N_j},\\
\mathbf{V}_{j} &:= \mathbf{H}_{j,j}\mathbf{Q}_{j}\mathbf{H}_{j,j}^{\cH}.\label{eq:matV}
\end{align}

At each iteration $n = 1,2,\ldots$, the block coordinate ascent solver amounts to updating the covariance matrices $\{\mathbf{Q}_{k}\}$ in a round robin fashion via (P4), where the solution obtained at the $(n-1)$-st iteration are exploited to compute the complex gradient~\eqref{eq:gradD}. The term $\Tr\left\{\mathbf{D}_{k}^{\cH}\mathbf{Q}_{k}\right\}$ discourages a ``selfish'' behavior of the $k$-th CR-to-CR link, which would otherwise try to simply minimize its own MSE, as in the game-theoretic formulations of \cite{Scutari} and \cite{WangSP10}. In the next subsection, the robust interference constraint will be translated to a tractable form, and (P4) will be re-stated accordingly.

% ----------------------------
\subsection{Equivalent robust interference constraint}
% ----------------------------
Constraint~\eqref{P5RobustInterf} renders (P4) a \emph{semi-infinite} program (cf.~\cite[Ch.~3]{Bertsekas}). An equivalent constraint in linear matrix inequality (LMI)
form will be derived next, thus turning (P4) into an equivalent \emph{semi-definite} program (SDP),
which can be efficiently solved in polynomial time by standard interior point methods.
To this end, the following lemma is needed.
\begin{lemma}
\label{lemma:S-Procedure} (S-Procedure \cite[p.~655]{Boyd}) Consider
$\mathbf{A},\mathbf{D} \in \mathbb{H}^{n \times n}, \mathbf{b} \in
\mathbb{C}^{n}, c, e \in \mathbb{R}$, and assume the interior condition holds, i.e.,
there exists an $\bar{\mathbf{x}}$ satisfying
$\bar{\mathbf{x}}^{\cH}\mathbf{D}\bar{\mathbf{x}} < e$. Then, the inequality
\begin{align}
\mathbf{x}^{\cH}\mathbf{A}\mathbf{x}+2\Re(\mathbf{b}^{\cH}\mathbf{x})+c \ge 0,~
\forall~ \mathbf{x}^{\cH}\mathbf{D}\mathbf{x} \le e
\end{align}
holds if and only if there exists $\theta \ge 0$ such that
\begin{align}
\left[\begin{array}{cc}
\theta \mathbf{D}+\mathbf{A} & \mathbf{b}\\
\mathbf{b}^{\cH} &c-e\theta
\end{array}\right] \succeq \mathbf{0} \, .
\end{align}
\end{lemma}
Using Lemma~\ref{lemma:S-Procedure}, the robust constraint~\eqref{P5RobustInterf} can be
equivalently reformulated as follows.
\begin{proposition}
\label{prop:RIC}
There exists $\theta_{k} \ge 0$, so that the robust interference
constraint (\ref{P5RobustInterf}) is equivalent to the following LMI
\begin{align}
\left[\begin{array}{cc}
\theta_{k}\mathbf{I}_{L\times M_k} -\left(\mathbf{I}_{L}\otimes \mathbf{Q}_{k}\right)
& - \textnormal{\Vect}(\mathbf{Q}_{k}^{\cH}\widehat{\mathbf{G}}_{k}^{\cH})\\
- \textnormal{\Vect}(\mathbf{Q}_{k}^{\cH}\widehat{\mathbf{G}}_{k}^{\cH})^{\cH} &\begin{array}{c}
\iota^{\textrm{max}}_{k}-\epsilon_k^2\theta_{k} \\
-\textnormal{\Tr}\{\widehat{\mathbf{G}}_{k}\mathbf{Q}_{k}\widehat{\mathbf{G}}^{\cH}_{k}\}\end{array}
\end{array}\right] \succeq \mathbf{0} \, .
\label{eq:robust_constraint}
\end{align}
\end{proposition}
\begin{IEEEproof}
Using the properties of the trace operator
$\Tr(\mathbf{Z}^{\cH} \mathbf{A}\mathbf{Z})=\Vect(\mathbf{Z})^{\cH}(\mathbf{I}\otimes\mathbf{A})\Vect(\mathbf{Z})$
and $\Tr(\mathbf{B}^{\cH}\mathbf{Z})=\Vect(\mathbf{B})^{\cH}\Vect(\mathbf{Z})$,
%$\forall~\mathbf{Z} \in \mathbb{C}^{M \times N}, \mathbf{A} \in \mathbb{C}^{M \times M}$,
constraint (\ref{P5RobustInterf}) can be re-written as
\begin{align}~\label{vect_equiv_constr}
-\mathbf{g}_{k}^{\cH}\left(\mathbf{I}_{L}\otimes \mathbf{Q}_{k}\right)\mathbf{g}_{k} &-
2\Re\left(\Vect(\mathbf{Q}_{k}^{\cH}\widehat{\mathbf{G}}_{k}^{\cH})^{\cH}\mathbf{g}_{k}\right) \nonumber \\
&\hspace{-2cm} + \iota^{\textrm{max}}_{k}-\Tr\{\widehat{\mathbf{G}}_{k}\mathbf{Q}_{k}\widehat{\mathbf{G}}^{\cH}_{k}\} \ge 0,
\, \forall~\|\mathbf{g}_{k}\|_{2} \le \epsilon_{k}
\end{align}
where $\mathbf{g}_{k} := \Vect(\Delta\mathbf{G}_{k}^{\cH})$.
Then, applying Lemma~\ref{lemma:S-Procedure} to~\eqref{vect_equiv_constr} yields readily~\eqref{eq:robust_constraint}.
\end{IEEEproof}

\begin{proposition}
Problem (P4) can be equivalently re-written as the following SDP form:
\begin{subequations}
\begin{align}
\textnormal{(P5)}\quad \mathop{\textnormal{\min}}_{\substack{\mathbf{Q}_{k}\succeq \mathbf{0}\\\mathbf{T}, \theta_{k} \ge 0}}\quad
&\textnormal{\Tr}\left\{\mathbf{T}\right\}-\textnormal{\Tr}\left\{\mathbf{D}_{k}^{\cH}\mathbf{Q}_{k}\right\}\label{P6ObjFunc}\\
\textnormal{\st} \quad \, & \textnormal{\Tr}\{\mathbf{Q}_{k}\} \le p^{\textrm{max}}_{k}\label{P6TxPower}\\
& \hspace{-1cm}\left[\begin{array}{cc}
\mathbf{H}_{k,k}\mathbf{Q}_{k}\mathbf{H}^{\cH}_{k,k}+\mathbf{R}_{k,k} &\mathbf{R}^{1/2}_{k,k}\\
\mathbf{R}^{1/2}_{k,k} &\mathbf{T}\end{array}\right] \succeq \mathbf{0}\label{P6SchurSDP}\\
& \hspace{-1.75cm} \left[\begin{array}{cc}
\theta_{k}\mathbf{I}_{L\times M_k} -\left(\mathbf{I}_{L}\otimes \mathbf{Q}_{k}\right) & - \textnormal{\Vect}(\mathbf{Q}_{k}^{\cH}\widehat{\mathbf{G}}_{k}^{\cH})\\
- \textnormal{\Vect}(\mathbf{Q}_{k}^{\cH}\widehat{\mathbf{G}}_{k}^{\cH})^{\cH} & \begin{array}{c}\iota^{\textrm{max}}_{k}-\epsilon_k^2\theta_{k} \\
- \textnormal{\Tr}\{\widehat{\mathbf{G}}_{k}\mathbf{Q}_{k}\widehat{\mathbf{G}}^{\cH}_{k}\}\end{array}
\end{array}\right] \succeq \mathbf{0} \, .
\label{P6RobustInterf}
\end{align}
\end{subequations}
\end{proposition}
\begin{IEEEproof}
First, note that~[cf.~\eqref{per-utility}]
\begin{align*}
u_k\left(\mathbf{Q}_{k},\mathbf{Q}_{-k}\right) =
\Tr\left\{\mathbf{I}_{N_k}-\mathbf{R}_{k,k}\left(\mathbf{H}_{k,k}\mathbf{Q}_{k}\mathbf{H}^{\cH}_{k,k}+\mathbf{R}_{k,k}\right)^{-1}\right\}.
\end{align*}
Thus, (P4) is equivalent to
 \begin{align*}
&\mathop{\min}_{\mathbf{Q}_{k}}\quad \Tr\left\{\mathbf{R}_{k,k}^{1/2}\left(\mathbf{H}_{k,k}\mathbf{Q}_{k}\mathbf{H}^{\cH}_{k,k}+\mathbf{R}_{k,k}\right)^{-1}\mathbf{R}_{k,k}^{1/2}\right\}\nonumber \\
&\hspace{4.5cm} -\Tr\left\{\mathbf{D}_{k}^{\cH}\mathbf{Q}_{k}\right\} \\
&\st \quad \eqref{P5PSDCsrt}-\eqref{P5RobustInterf}.
 \end{align*}
Then, an auxiliary matrix variable $\mathbf{Y}$ is introduced such that
$\mathbf{Y} \succeq \mathbf{R}_{k,k}^{1/2}\left(\mathbf{H}_{k,k}\mathbf{Q}_{k}\mathbf{H}^{\cH}_{k,k}+\mathbf{R}_{k,k}\right)^{-1}\mathbf{R}_{k,k}^{1/2}$,
which can be equivalently recast as~\eqref{P6SchurSDP} by using the Schur complement~\cite{Boyd}. Combining the LMI form of the robust
interference constraint~\eqref{eq:robust_constraint}, the formulation of (P5) follows immediately.
\end{IEEEproof}

Problem (P3) can be solved in a \emph{centralized} fashion
upon collecting CR-to-CR channels $\{\mathbf{H}_{j,k}\}$,
CR-to-PU estimated channels $\{\widehat{\mathbf{G}}_{k}\}$,
and confidence intervals $\{\epsilon_k\}$ at a CR fusion center.
The optimal transmit-covariance matrices can be found at the
fusion center by solving (P5), and sent back to all CRs.
This centralized scheme is tabulated as Algorithm~\ref{algo:iterSDP_centralized},
where $\mathbf{Q}_{k}^{(n)}$ denotes the transmit-covariance matrix of
CR $U_k^t$ at iteration $n$ of the block coordinate ascent algorithm;
$\mathbf{Q}^{(n)}:=\left(\mathbf{Q}_{1}^{(n)},\ldots,\mathbf{Q}_{K}^{(n)}\right)$ represents the set of transmit-covariance matrices at iteration $n$;
$\cU(\cdot)$ is the objective function~\eqref{P4ObjFunc}.
A simple stopping criterion for terminating the iterations is
$\cU\left(\mathbf{Q}^{(n)}\right) - \cU\left(\mathbf{Q}^{(n-1)}\right)< \upsilon$,
where $\upsilon > 0$ denotes a preselected threshold.

%%%%%%%%%%%%%%%%%%%%%%%%%%%%%%%%%%%%%%%%%%%%%%%%%%%%
\begin{algorithm}[t]
\caption{Centralized robust sum-MSE minimization}
\label{algo:iterSDP_centralized}
\begin{algorithmic}[1]
\State Collect all channel matrices $\{\mathbf{H}_{j,k}\}$, and noise powers $\{\sigma_k^2\}$
\State Collect all CR-to-PU channel matrices $\{\widehat{\mathbf{G}}_{k}\}$, and confidence intervals $\{\epsilon_k\}$
\State Initialize $\mathbf{Q}_{k}^{(0)} =\mathbf{0}, \forall~ k \in \cK$
\Repeat \quad ($n = 1,2,\ldots$)
\For{$k=1,2,\dots,K$}
    \State Compute $\mathbf{D}_{k}^{(n)}$ via~\eqref{eq:gradD}
    \State Update $\mathbf{Q}_{k}^{(n)}$ by solving (P5) [(P6) for the proximal point-based method]
\EndFor
\Until $\cU\left(\mathbf{Q}^{(n)}\right) - \cU\left(\mathbf{Q}^{(n-1)}\right)< \upsilon$
\State Calculate $\{\mathbf{W}^{\textrm{opt}}_{k}\}$ via~\eqref{optW}
\State Broadcast optimal transmit- and receive-beamformers
\end{algorithmic}
\end{algorithm}
%%%%%%%%%%%%%%%%%%%%%%%%%%%%%%%%%%%%%%%%%%%%%%%%%%%%%%

To alleviate the high communication cost associated with the centralized setup,
and ensure scalability with regards to network size and enhanced robustness to fusion center failure,
a distributed optimization algorithm is generally desirable. It can be noticed that the proposed coordinate ascent approach lends itself to a
\emph{distributed} optimization procedure that can be implemented in an \emph{on-line} fashion. Specifically, each CR $U_k^t$ can update locally $\mathbf{Q}_{k}$ via (P5) based on
a measurement of the interference $\mathbf{R}_{k,k}^{(n)}$~\cite{SJK}, and the following information necessary to compute the complex gradient~\eqref{eq:gradD}:
\emph{i)} its covariance matrix $\mathbf{Q}_{k}^{(n-1)}$ obtained at the previous iteration;
\emph{ii)} matrices $\{\mathbf{B}_j^{(n)}\}$ and $\{\mathbf{V}_j^{(n)}\}$ obtained from the neighboring CR links via local message passing.
Furthermore, it is clear that the terms in~\eqref{eq:gradD} corresponding to CRs located far away from CR $U_k^t$ are negligible due to the path loss effect in channel $\{\mathbf{H}_{j,k}\}_{j \neq k}$;
hence, summation in~\eqref{eq:gradD} is only limited to the interfering CRs, and consequently, matrices $\{\mathbf{B}_j^{(n)}\}$ and $\{\mathbf{V}_j^{(n)}\}$ need to be exchanged only locally.
The overall distributed scheme is tabulated as Algorithm~\ref{algo:iterSDP}.
The on-line implementation of the iterative optimization allows tracking of slow variations of the channel matrices;
in this case, cross-channels $\{\mathbf{H}_{j,k}\}$ in Algorithm~\ref{algo:iterSDP} need to be re-acquired whenever a change is detected.
Finally, notice that instead of updating the transmit-covariances in a Gauss-Seidel fashion, Jacobi iterations or asynchronous schemes~\cite{Bertsekas97} can be alternatively employed.

%%%%%%%%%%%%%%%%%%%%%%%%%%%%%%%%%%%%%%%%%%%%%%%%%%%%
\begin{algorithm}[t]
\caption{Distributed on-line robust sum-MSE minimization}
\label{algo:iterSDP}
\begin{algorithmic}[1]
\State Initialize $\mathbf{Q}_{k}^{(0)}=\mathbf{0}, \forall~k \in \cK$
\Repeat \quad ($n = 1,2,\ldots$)
\For{$k=1,2,\dots,K$}
    \State $U_k^t$ acquires $\mathbf{H}_{j,k}$ from its neighboring $U_j^r$
		\State transmit $\{\mathbf{B}_{k}^{(n)},\mathbf{V}_{k}^{(n)}\}$ to neighboring nodes
		\State receive $\{\mathbf{B}_{j}^{(n)},\mathbf{V}_{j}^{(n)}\}_{j \neq k}$ from neighboring nodes
    \State Compute $\mathbf{D}_{k}^{(n)}$ via~\eqref{eq:gradD}
    \State Measure $\mathbf{R}_{k,k}^{(n)}$
    \State Update $\mathbf{Q}_{k}^{(n)}$ by solving (P5) [(P6) for the proximal point-based method]
    \State Update $\mathbf{W}_{k}^{(n)}$ via~\eqref{optW}
    \State Transmit and receive signals using $\mathbf{Q}_{k}^{(n)}$ and $\mathbf{W}_{k}^{(n)}$
\EndFor
\Until $u_k\left(\mathbf{Q}^{(n)}\right) - u_k\left(\mathbf{Q}^{(n-1)}\right) < \upsilon$, $\forall~k \in \cK$
\end{algorithmic}
\end{algorithm}
%%%%%%%%%%%%%%%%%%%%%%%%%%%%%%%%%%%%%%%%%%%%%%%%%%%%%%

\noindent \textbf{Remark 5} \textit{(Multiple PU receivers).}
For ease of exposition, the formulated robust optimization problems consider a single PU receiver. Clearly, in case of $N_{\textrm{PU}} >1$ receiving PU devices, or when a grid of $N_{\textrm{PU}}$ potential PU locations is obtained from the sensing phase~\cite{EmiSJGG11}, a robust interference constraint for each of the $K N_{\textrm{PU}}$ CR-to-PU links must be included in (P3). As for (P5),
it is still an SDP, but with $N_{\textrm{PU}}$ LMI constraints~\eqref{P6RobustInterf}, and one additional optimization variable ($\theta_k$) per PU receiver.

\noindent \textbf{Remark 6} \textit{(Network synchronization).}
Similar to~\cite{Ding,Cioffi-twc08,Meisam,Shi,Gharavol-tvt10,
Rui08,Liangjstsp08,SJK,Scutari,WangSP10,Zheng,PalomarCL03,Liang,KhasibSL11,Chiang-tsp11,Poor09},
time synchronization is assumed to have been acquired. In practice, accurate time synchronization
among the CR transmitters can be attained (and maintained during operation)
using e.g., pairwise broadcast synchronization protocols~\cite{Serpedin08}, consensus-based methods~\cite{Zennaro11}, or mutual network synchronization approaches~\cite{Kunz08}. To this end, CRs have to exchange synchronization beacons on a regular basis; clearly, the number of time slots occupied by the transmission of these beacons depends on the particular algorithm implemented, the CR network size, and the targeted synchronization accuracy. For example, the algorithm in~\cite{Serpedin08} entails two message exchanges per transmitter pairs, while the message-passing overhead of consensus-based methods generally depends on the wanted synchronization accuracy~\cite{Zennaro11}. Since the CR network operates in an underlay setup, this additional message passing can be performed over the primary channel(s). Alternatively, a CR control channel can be employed to avoid possible synchronization errors due to the interference inflicted by the active PU transmitters. Analyzing the effect of mistiming constitutes an interesting research direction, but it goes beyond the scope and page limit of this paper.

% ----------------------------
\subsection{Convergence}
% ----------------------------
Since the original optimization problem (P3) is non-convex, convergence of the
block coordinate ascent with local convex approximation has to be analytically established.
To this end, recall that (P4) and (P5) are equivalent; thus,
convergence can be asserted by supposing that (P4) is solved per
Gauss-Seidel iteration instead of (P5).
The following lemma (proved in Appendix~\ref{Appendix:convexfunc}) is needed first.
\begin{lemma}
For each $k \in \cK$, the feasible set of problem (P4), namely
$\mathcal{Q}_{k} := \{\mathbf{Q}_{k}|\mathbf{Q}_{k} \in
(\ref{P5PSDCsrt})-(\ref{P5RobustInterf})\}$, is convex. The
real-valued function $f_{k}(\mathbf{Q}_{k},\mathbf{Q}_{-k})$ is
convex in $\mathbf{Q}_{k}$ over the feasible set $\mathcal{Q}_{k}$,
when the set $\mathbf{Q}_{-k} := \{\mathbf{Q}_{j},j\neq k\}$ is fixed.
\label{lemma:convexfunc}
\end{lemma}
%\begin{IEEEproof}
%
%\end{IEEEproof}

Based on Lemma~\ref{lemma:convexfunc}, convergence of the block coordinate ascent algorithm is established next.

\begin{proposition}
\label{prop:objconverge}
The sequence of objective function values~\eqref{P4ObjFunc} obtained by the coordinate
ascent algorithm with concave-convex procedure converges.
\end{proposition}
\begin{IEEEproof}
It suffices to show that the sequence of objective values~\eqref{P4ObjFunc} is monotonically non-decreasing.
Since the objective function value is bounded from above,
the function value sequence must be convergent by invoking the monotone convergence theorem.
Letting $\widetilde{\cU}_{k}(\cdot)$ denote the objective function~\eqref{P5ObjFunc}, which is
the concave surrogate of $\cU(\cdot)$ as the original objective~\eqref{P4ObjFunc}, consider
\begin{align}
&\mathbf{Q}_{k}^{(n)} := \nonumber \\
&\argmax_{\mathbf{Q}_{k} \in \cQ_{k}}\,\,
\widetilde{\cU}_{k}\left(\mathbf{Q}_{k};\mathbf{Q}_{1}^{(n)},\ldots,\mathbf{Q}_{k-1}^{(n)}, \mathbf{Q}_{k+1}^{(n-1)},\ldots,\mathbf{Q}_{K}^{(n-1)}\right)
\end{align}
where $n$ stands for the iteration index. Furthermore, define
\begin{align}
\mathbf{Z}_{k}^{(n)} &:= (\mathbf{Q}_{1}^{(n+1)},\ldots,\mathbf{Q}_{k}^{(n+1)}, \mathbf{Q}_{k+1}^{(n)},\ldots,\mathbf{Q}_{K}^{(n)}), \\
\tilde{\mathbf{Q}}^{(n)}_{-k} &:= (\mathbf{Q}_{1}^{(n+1)},\ldots,\mathbf{Q}_{k-1}^{(n+1)},
\mathbf{Q}_{k+1}^{(n)},\ldots,\mathbf{Q}_{K}^{(n)}).~\label{eq:Qminusk}
\end{align}
Then, for all $k \in \cK$, it holds that
%\begin{small}
\begin{subequations}
\begin{align}
\hspace{-0.2cm} \cU\left(\mathbf{Z}_{k}^{(n)}\right)
&= u_k\left(\mathbf{Q}_{k}^{(n+1)},\tilde{\mathbf{Q}}^{(n)}_{-k}\right) +
f_k\left(\mathbf{Q}_{k}^{(n+1)},\tilde{\mathbf{Q}}^{(n)}_{-k}\right) \\
& \ge u_k\left(\mathbf{Q}_{k}^{(n+1)},\tilde{\mathbf{Q}}^{(n)}_{-k}\right) +
f_k\left(\mathbf{Q}_{k}^{(n)},\tilde{\mathbf{Q}}^{(n)}_{-k}\right)\nonumber \\
& \hspace{1.6cm}+ \Tr\left\{\mathbf{D}^{\cH}_{k}\left(\mathbf{Q}_{k}^{(n+1)}
- \mathbf{Q}_{k}^{(n)}\right)\right\} \label{ConvexPropt} \\
& \ge u_k\left(\mathbf{Q}_{k}^{(n)},\tilde{\mathbf{Q}}^{(n)}_{-k}\right) +
f_k\left(\mathbf{Q}_{k}^{(n)},\tilde{\mathbf{Q}}^{(n)}_{-k}\right)\nonumber \\
& \hspace{1.6cm}+ \Tr\left\{\mathbf{D}^{\cH}_{k}\left(\mathbf{Q}_{k}^{(n)}
- \mathbf{Q}_{k}^{(n)}\right)\right\} \label{OptSolu} \\
& = \cU\left(\mathbf{Z}_{k-1}^{(n)}\right)
\end{align}
\end{subequations}
%\end{small}
where~(\ref{ConvexPropt}) follows from the convexity of $f_k(\cdot)$ established in Lemma~\ref{lemma:convexfunc}; ~\eqref{OptSolu} holds because $\mathbf{Q}_{k}^{(n+1)}$ is the optimal solution of (P4) for fixed $\tilde{\mathbf{Q}}^{(n)}_{-k}$.

To complete the proof, it suffices to show that $\cU\left(\mathbf{Q}^{(n+1)}\right)$ is
monotonically non-decreasing, namely that
\begin{align}
\cU\left(\mathbf{Q}^{(n+1)}\right)
\ge \cU\left(\mathbf{Z}_{K-1}^{(n)}\right)
\ge \ldots
\ge \cU\left(\mathbf{Z}_{1}^{(n)}\right)
\ge \cU\left(\mathbf{Q}^{(n)}\right)\label{monotonU}
\end{align}
\end{IEEEproof}

Interestingly, by inspecting the structure of $\{\mathbf{H}_{k,k}, k \in \cK \}$,
it is also possible to show that every limit point generated by the coordinate ascent
algorithm with local convex approximation satisfies the first-order optimality conditions.
Conditions on $\{\mathbf{H}_{k,k}, k \in \cK \}$ that guarantee stationarity of the
limit points are provided next. First, it is useful to establish strict concavity
of the objective~\eqref{P5ObjFunc} in the following lemma
proved in Appendix~\ref{Appendix:strictconcave}.
\begin{lemma}
If the channel matrices $\{\mathbf{H}_{k,k}, k \in \cK \}$ of the CR links
$\{U_k^t \rightarrow U_k^r\}$ have full column rank,
then the objective function~\eqref{P5ObjFunc} is strictly concave in $\mathbf{Q}_{k}$.
\label{lemma:strictconcave}
\end{lemma}

We are now ready to establish stationarity of the limit points.
\begin{theorem}
If matrices $\{\mathbf{H}_{k,k}, k \in \cK \}$ have full column rank,
then every limit point of the coordinate ascent algorithm with concave-convex procedure
is a stationary point of (P3). \label{theoremKKT}
\end{theorem}
\begin{IEEEproof}
The proof of Theorem~\ref{theoremKKT} relies on the basic convergence claim
of the block coordinate descent method in~\cite[Ch.~2]{Bertsekas} and \cite{Meisam}.
What must be shown is that every limit point of the algorithm satisfies the
first-order optimality conditions over the Cartesian product of the closed convex sets.
Let $\bar{\mathbf{Q}} := \left(\bar{\mathbf{Q}}_{1}, \ldots, \bar{\mathbf{Q}}_{K}\right)$
be a limit point of the sequence $\{\mathbf{Q}^{(n)}\}$, and $\{\mathbf{Q}^{(n_j)}|j=1,2,\ldots\}$
a subsequence that converges to $\bar{\mathbf{Q}}$. First, we will show that
$\lim\limits_{j\rightarrow\infty}\mathbf{Q}_{1}^{(n_j+1)}=\bar{\mathbf{Q}}_{1}$.
Argue by contradiction, i.e., assume that $\{\mathbf{Q}_{1}^{(n_{j}+1)}-\mathbf{Q}_{1}^{(n_j)}\}$ does not converge to zero.
Define $\gamma^{(n_j)} := \|\mathbf{Q}_{1}^{(n_{j}+1)}-\mathbf{Q}_{1}^{(n_j)}\|_{F}$.
By possibly restricting to a subsequence of $\{n_j\}$, it follows that there exists some $\bar{\gamma}>0$
such that $\bar{\gamma}\le \gamma^{(n_j)}$ for all $j$. Let
$\mathbf{S}_{1}^{(n_j)} :=(\mathbf{Q}_{1}^{(n_{j}+1)}-\mathbf{Q}_{1}^{(n_{j})})/\gamma^{(n_j)}$. Thus, we have that
$\mathbf{Q}_{1}^{(n_j+1)} =\mathbf{Q}_{1}^{(n_j)}+\gamma^{(n_j)}\mathbf{S}_{1}^{(n_j)}$
and $\|\mathbf{S}_{1}^{(n_j)}\|_{F}=1$.
Because $\mathbf{S}_{1}^{(n_j)}$ belongs to a compact set,
it can be assumed convergent to a limit point $\bar{\mathbf{S}}_{1}$
along with a subsequence of $\{n_j\}$.

Since it holds that $0 \le \epsilon\bar{\gamma}\le \gamma^{(n_j)}$ for all $\epsilon \in [0,1]$,
the point $\mathbf{Q}_{1}^{(n_j)}+\epsilon\bar{\gamma}\mathbf{S}_{1}^{(n_j)}$ lies on the segment
connecting two feasible points $\mathbf{Q}_{1}^{(n_j)}$ and $\mathbf{Q}_{1}^{(n_j+1)}$.
Thus, $\mathbf{Q}_{1}^{(n_j)}+\epsilon\bar{\gamma}\mathbf{S}_{1}^{(n_j)}$
is also feasible due to the convexity of $\cQ_1$ [cf.~Lemma~\ref{lemma:convexfunc}].
Moreover, concavity of $\widetilde{\cU}_1(\cdot; \mathbf{Q}_{-1}^{(n_j)})$ implies that $\widetilde{\cU}_1$ is monotonically non-decreasing
in the interval connecting point $\mathbf{Q}_1^{(n_j)}$ to $\mathbf{Q}_{1}^{(n_j+1)}$ over the set $\cQ_1$.
Hence, it readily follows that
\begin{align}
\widetilde{\cU}_1(\mathbf{Q}_{1}^{(n_j+1)}; \mathbf{Q}_{-1}^{(n_j)})\ge
\widetilde{\cU}_1(\mathbf{Q}_{1}^{(n_j)}+\epsilon\bar{\gamma}\mathbf{S}_{1}^{(n_j)}; \mathbf{Q}_{-1}^{(n_j)})\nonumber \\
\ge \widetilde{\cU}_1(\mathbf{Q}_{1}^{(n_j)}; \mathbf{Q}_{-1}^{(n_j)}).~\label{KKTMonoton}
\end{align}

Note that $\widetilde{\cU}_1(\cdot)$ is a tight lower bound of $\cU(\cdot)$ at each current feasible point. Also, from~\eqref{monotonU}, $\widetilde{\cU}_1(\mathbf{Q}_{1}^{(n_j+1)}; \mathbf{Q}_{-1}^{(n_j)})$
is guaranteed to converge to $\widetilde{\cU}_1(\bar{\mathbf{Q}})$ as $j\rightarrow\infty$.
Thus, upon taking the limit as $j\rightarrow\infty$ in \eqref{KKTMonoton}, it follows that
\begin{align}
\widetilde{\cU}_1(\bar{\mathbf{Q}}_{1}+\epsilon\bar{\gamma}\bar{\mathbf{S}}_{1}; \bar{\mathbf{Q}}_{-1})=
\widetilde{\cU}_1(\bar{\mathbf{Q}}), \quad \forall~\epsilon \in [0,1] \, .~\label{KKTMultiOpt}
\end{align}
However, since $\bar{\gamma}\bar{\mathbf{S}}_{1}\neq \mathbf{0}$, \eqref{KKTMultiOpt} contradicts the unique maximum condition implied by the strict concavity of $\widetilde{\cU}_1(\cdot;\cdot)$ in $\mathbf{Q}_{1}$~[cf.~Lemma~\ref{lemma:strictconcave}].
Therefore, $\mathbf{Q}_{1}^{(n_j+1)}$ converges to $\bar{\mathbf{Q}}_{1}$ as well.

Consider now checking the optimality condition for $\bar{\mathbf{Q}}_{1}$.
Since $\mathbf{Q}_{1}^{(n_j+1)}$ is the local (and also global) maximum of
$\widetilde{\cU}_1(\cdot; \mathbf{Q}_{-1}^{(n_j)})$, we have that
\begin{align}
\Re\left\{\Tr\left\{\nabla_1\widetilde{\cU}_1\left(\mathbf{Q}_{1}^{(n_j+1)};\mathbf{Q}_{-1}^{(n_j)}\right)^{\cH}
\left(\mathbf{Q}_{1}-\mathbf{Q}_{1}^{(n_j+1)}\right)\right\}\right\} &\le 0, \nonumber \\
& \hspace{-2.5cm} \forall~\mathbf{Q}_{1} \in \cQ_1~\label{KKTOptCond1}
\end{align}
where $\nabla_1\widetilde{\cU}_1(\cdot)$ denotes the gradient of $\widetilde{\cU}_1(\cdot)$ with respect to $\mathbf{Q}_{1}$.
Taking the limit as $j\rightarrow\infty$, and using the fact that
$\nabla_1\widetilde{\cU}_1(\bar{\mathbf{Q}})=\nabla_1\cU_1(\bar{\mathbf{Q}})$,
it is easy to show that
\begin{align}
\Re\left\{\Tr\left\{\nabla_1{\cU}(\bar{\mathbf{Q}})^{\cH}
(\mathbf{Q}_{1}-\bar{\mathbf{Q}}_{1})\right\}\right\}
\le 0, \quad \forall~\mathbf{Q}_{1} \in \cQ_1.~\label{KKTOptCond1_limit}
\end{align}
Using similar arguments, it holds that
\begin{align}
\Re\left\{\Tr\left\{\nabla_i{\cU}(\bar{\mathbf{Q}})^{\cH}
(\mathbf{Q}_{i}-\bar{\mathbf{Q}}_{i})\right\}\right\} &\le 0, \nonumber \\
& \hspace{-1.5cm} \forall~\mathbf{Q}_{i} \in \cQ_i,~i=1,2,\ldots,K~\label{KKTOptCond_limits}
\end{align}
which establishes the stationarity of $\bar{\mathbf{Q}}$ and completes the proof.
\end{IEEEproof}

% ----------------------------
\subsection{Proximal point-based robust algorithm}
\label{sec:proximal}
% ----------------------------
The full column rank requirement can be quite restrictive in practice; e.g., if $M_k > N_k$ for at least one CR link,
or in the presence of spatially correlated MIMO channels~\cite{Kermoal02}. Furthermore, computing the rank of channel matrices increases the computational burden to an extent that may not be affordable by the CRs. In this section, an alternative approach based on proximal-point regularization~\cite{Rockf76} is pursued to ensure convergence, without requiring restrictions on the antenna configuration and the channel rank.

The idea consists in penalizing the objective of (P4) using a quadratic regularization term $\frac{1}{2\tau_k}\|\mathbf{Q}_{k}-\mathbf{Q}_k^{(n-1)}\|^{2}_{F}$,
with a given sequence of numbers $\tau_k > 0$. Then, (P5) is modified as
\begin{subequations}
\begin{align}
\text{(P6)}\,\, \mathop{\min}_{\substack{\mathbf{Q}_{k}\succeq \mathbf{0},\theta_{k} \ge 0 \\ \mathbf{T},\mathbf{Y}}}\,
&\Tr\left\{\mathbf{T}\right\}-\Tr\left\{\mathbf{D}_{k}^{\cH}\mathbf{Q}_{k}\right\}
+ \frac{1}{2\tau_k}\Tr\{\mathbf{Y}\}\label{P7ObjFunc}\\
\st \quad
&\left[\begin{array}{cc}
\mathbf{I}_{M_k} &\mathbf{Q}_k-\mathbf{Q}_k^{(n-1)}\\
\mathbf{Q}_k- \mathbf{Q}_k^{(n-1)} &\mathbf{Y}\end{array}\right] \succeq \mathbf{0}\label{P7SchurSDP2}\\
&\,\,\eqref{P6TxPower},\eqref{P6SchurSDP},\eqref{P6RobustInterf}\nonumber
\end{align}
\end{subequations}
where~\eqref{P7SchurSDP2} is derived by using the Schur complement through the auxiliary variable $\mathbf{Y}$.

The role of $\frac{1}{2\tau_k} \|\mathbf{Q}_{k}-\mathbf{Q}_k^{(n-1)}\|^{2}_{F}$ is to render the cost in~\eqref{P7ObjFunc} strictly convex and coercive.
Moreover, for small values of $\tau_k$, the optimization variable $\mathbf{Q}_{k}$ is forced to stay ``close'' to $\mathbf{Q}_{k}^{(n-1)}$ obtained at the previous iteration,
thereby improving the stability of the iterates~\cite[Ch.~6]{Bertsekas09}.
Centralized and distributed schemes with the proximal point regularization are given by Algorithms~\ref{algo:iterSDP_centralized} and~\ref{algo:iterSDP}, respectively, with problem (P6) replacing (P5).
Convergence of the resulting schemes is established in the following theorem. To avoid ambiguity, these proximal point-based algorithms will be hereafter referred as Algorithms~\ref{algo:iterSDP_centralized}(P) and~\ref{algo:iterSDP}(P), respectively.

\begin{theorem}
Suppose that the sequence $\{\mathbf{Q}^{(n)}\}$ generated by Algorithm~1(P) (Algorithm~2(P)) has a limit point.
Then, every limit point is a stationary point of (P3).
\label{PGS-theoremKKT}
\end{theorem}
\begin{IEEEproof}
The Gauss-Seidel method with a proximal point regularization converges
without any underlying convexity assumptions~\cite{Grippo}.
A modified version of the proof is reported here,
where the local convex approximation~\eqref{eq:lla} and
the peculiarities of the problem at hand are leveraged to
establish not only convergence of the algorithm, but also optimality of the obtained solution.

Assume there exists a subsequence $\{\mathbf{Q}^{(n_j)}|j=1,2,\ldots\}$ converging to a
limit point $\bar{\mathbf{Q}}:= \left(\bar{\mathbf{Q}}_{1}, \ldots, \bar{\mathbf{Q}}_{K}\right)$.
Let $\mathbf{Q}_{k}^{(n+1)}$ be obtained as
\begin{align}
&\mathbf{Q}_{k}^{(n+1)} := \nonumber \\
&\argmax_{\mathbf{Q}_{k} \in \cQ_{k}}\,\,
\widetilde{\cU}_{k}\left(\mathbf{Q}_{k};\mathbf{Q}_{1}^{(n+1)},\ldots,\mathbf{Q}_{k-1}^{(n+1)}, \mathbf{Q}_{k+1}^{(n)},\ldots,\mathbf{Q}_{K}^{(n)}\right)\nonumber \\
& \hspace{4.5cm} -\frac{1}{2\tau_k}\|\mathbf{Q}_{k}-\mathbf{Q}^{(n)}_k\|^2_{F}\, . \label{XgenrtRule}
\end{align}
Thus, it follows that~[cf.~\eqref{eq:Qminusk}]
\begin{align}
\widetilde{\cU}_1(\mathbf{Q}_{1}^{(n_j+1)}; \tilde{\mathbf{Q}}_{-1}^{(n_j)}) &\ge
\widetilde{\cU}_1(\mathbf{Q}_{1}^{(n_j)}; \tilde{\mathbf{Q}}_{-1}^{(n_j)}) \nonumber \\
& \hspace{0.7cm} + \frac{1}{2\tau_k}\|\mathbf{Q}_{1}^{(n_j+1)}-\mathbf{Q}_{1}^{(n_j)}\|^2_{F} \, .
\label{PGS-KKTMonoton}
\end{align}
Going along the lines of the proof of Theorem~\ref{theoremKKT}, it holds that
\begin{align}
\lim\limits_{j\rightarrow\infty}\widetilde{\cU}_1(\mathbf{Q}_{1}^{(n_j+1)}; \tilde{\mathbf{Q}}_{-1}^{(n_j)})=
\lim\limits_{j\rightarrow\infty}\widetilde{\cU}_1(\mathbf{Q}_{1}^{(n_j)}; \tilde{\mathbf{Q}}_{-1}^{(n_j)})
=\widetilde{\cU}_1(\bar{\mathbf{Q}}).~\label{PGS-KKTMultiOpt}
\end{align}
Therefore, taking the limit as $j\rightarrow\infty$ in \eqref{PGS-KKTMonoton}, one arrives at
\begin{align}
\lim\limits_{j\rightarrow\infty}\|\mathbf{Q}_{1}^{(n_j+1)}-\mathbf{Q}_{1}^{(n_j)}\|_{F}^2 = 0
\end{align}
which implies that $\mathbf{Q}_{1}^{(n_j+1)}$ also converges to $\bar{\mathbf{Q}}_{1}$.

Since $\mathbf{Q}_{1}^{(n_j+1)}$ is generated as in \eqref{XgenrtRule}, it satisfies the optimality condition
\begin{align}
\Re\Bigg\{
&\Tr\Bigg\{\left[\nabla_1\widetilde{\cU}_1(\mathbf{Q}_{1}^{(n_j+1)};\mathbf{Q}_{-1}^{(n_j)})
-\frac{1}{\tau_1}(\mathbf{Q}_{1}^{(n_j+1)}-\mathbf{Q}_{1}^{(n_j)})\right]^{\cH} \nonumber \\
&\hspace{1.4cm} (\mathbf{Q}_{1}-\mathbf{Q}_{1}^{(n_j+1)})\Bigg\}
\Bigg\}
\le 0, \quad \forall~\mathbf{Q}_{1} \in \cQ_1.~\label{PGS-KKTOptCond1}
\end{align}
Taking the limit as $j\rightarrow\infty$ in~\eqref{PGS-KKTOptCond1}, and using again the fact that
$\nabla_1\widetilde{\cU}_1(\bar{\mathbf{Q}})=\nabla_1\cU_1(\bar{\mathbf{Q}})$, we obtain
\begin{align}
\Re\left\{\Tr\left\{\nabla_1{\cU}(\bar{\mathbf{Q}})^{\cH}(\mathbf{Q}_{1}-\bar{\mathbf{Q}}_1)\right\}\right\}
\le 0, \quad \forall~\mathbf{Q}_{1} \in \cQ_1 \, .
\label{PGS-KKTOptCond1_limit}
\end{align}
Then, repeating the same argument for all~$k \in \cK$, leads to
\begin{align}
\Re\left\{\Tr\left\{\nabla_k{\cU}(\bar{\mathbf{Q}})^{\cH}(\mathbf{Q}_{k}-\bar{\mathbf{Q}}_k)\right\}\right\}
\le 0, \quad \forall~\mathbf{Q}_{k} \in \cQ_k~\label{PGS-KKTOptCond_limits}
\end{align}
which shows that the limit point $\bar{\mathbf{Q}}$ is also a stationary point.
\end{IEEEproof}

As asserted in Theorem~\ref{PGS-theoremKKT}, Algorithms~1(P) and~2(P) converge to a stationary point of (P3) for any possible antenna configuration. The price to pay however, is a possibly slower convergence rate that is common to proximal point-based methods~\cite[Ch.~6]{Bertsekas09} (see also the numerical tests in Section~\ref{sec:numericalresults}). For this reason, the proximal point-based method should be used in either a centralized or a distributed setup whenever the number of transmit-antennas exceeds that of receive-antennas in at least one transmitter-receiver pair. In this case, Algorithms~1(P) and~2(P) ensure first-order optimality of the solution obtained. When $M_k \leq N_k$, for all $k \in \cK$, the two solvers have complementary strengths in convergence rate and computational complexity. Specifically, Algorithms~1 and~2 require the rank of all CR direct channel matrices $\{\mathbf{H}_{k,k}\}$ beforehand, which can be computationally burdensome, especially for a high number of antenna elements. If the rank determination can be afforded, and the convergence rate is at a premium, then Algorithms~1 and~2 should be utilized.

\section{Aggregate Interference Constraints}\label{Section:PrimalDecom}
Suppose now that the individual interference budgets $\{\iota_{k}^{\textrm{max}}\}$ are not available a priori.
Then, the aggregate interference power $\{\iota^{\textrm{max}}\}$ has to be divided among transmit-CRs by the resource allocation scheme
in order for the overall system performance to be optimized.
Accordingly, (P3) is modified as follows to incorporate a robust constraint on the total interference power inflicted to the PU node:
\begin{subequations}
\begin{align}
\textrm{(P7)}\quad
\mathop{\max}\limits_{\{\mathbf{Q}_{k}\}^{K}_{k=1}} &\sum^{K}_{k=1}u_{k}(\{\mathbf{Q}_{k}\})\label{P8ObjFunc}\\
\st \quad &\textrm{Tr}\{\mathbf{Q}_{k}\} \le p^{\textrm{max}}_{k}, \, k \in \cK\label{P8TxPower}\\
& \hspace{-2cm} \sum_{k=1}^{K}\Tr\{\mathbf{G}_{k}\mathbf{Q}_{k}\mathbf{G}^{\cH}_{k}\} \le
\iota^{\textrm{max}}, \, \forall~ \Delta\mathbf{G}_{k} \in \mathcal{G}_{k}, \, k \in \cK. \label{CuplingRobustInterf}
\end{align}
\end{subequations}
The new interference constraint~\eqref{CuplingRobustInterf} couples the CR nodes (or, more precisely, the subset of transmit-CR nodes in the proximity of the PU receiver).
Thus, the overhead of message passing increases since cooperation among coupled CR nodes is needed.

A common technique for dealing with coupled constraints is the dual decomposition method~\cite{Bertsekas}, which facilitates evaluation of the dual function by dualizing the coupled constraints.
However, since (P7) is non-convex and non-separable, the duality gap is generally non-zero. Thus, the primal variables obtained during the intermediate iterates may not be feasible, i.e.,
transmit-covariances can possibly lead to violation of the interference constraint. Since the ultimate goal is to design an on-line algorithm where~\eqref{CuplingRobustInterf} must be satisfied during network operation,
the primal decomposition technique is well motivated to cope with the coupled interference constraints~\cite{SJK}.
To this end, consider introducing two sets of auxiliary variables $\{\iota_{k}\}$ and $\{t_k\}$ in problem (P7), which is equivalently re-formulated as
\begin{subequations}
\begin{align}
\textrm{(P8)}\nonumber \\
\mathop{\max}\limits_{\{\mathbf{Q}_{k},\iota_k,t_k\}}
&\sum^{K}_{k=1}\textrm{Tr}\left\{\mathbf{H}_{k,k}\mathbf{Q}_{k}\mathbf{H}^{\cH}_{k,k}
\left(\mathbf{H}_{k,k}\mathbf{Q}_{k}\mathbf{H}^{\cH}_{k,k}+\mathbf{R}_{k,k}\right)^{-1}\right\}\label{P9ObjFunc}\\
\st \quad &\textrm{Tr}\{\mathbf{Q}_{k}\} \le p^{\textrm{max}}_{k}, \quad k \in \cK~\label{P9TxPower}\\
& \textrm{Tr}\{\mathbf{G}_{k}\mathbf{Q}_{k}\mathbf{G}^{\cH}_{k}\} \le t_{k},~
\forall~\Delta\mathbf{G}_{k} \in \mathcal{G}_{k}, k \in \cK~\label{P9RobustInterf}\\
& 0 \le t_k \le \iota_{k},\quad k \in \cK \\
& \sum_{k = 1}^{K} \iota_{k} \le \iota^{\textrm{max}}\, .
\end{align}
\end{subequations}
For fixed $\{\iota_{k}\}$, the inner maximization subproblem turns out to be
\begin{subequations}
\begin{align}
\textrm{(P9)}\quad p(&\{\iota_{k}\}) := \nonumber \\
\mathop{\max}\limits_{\{\mathbf{Q}_{k},t_k\}}
&\sum^{K}_{k=1}\textrm{Tr}\left\{\mathbf{H}_{k,k}\mathbf{Q}_{k}\mathbf{H}^{\cH}_{k,k}
\left(\mathbf{H}_{k,k}\mathbf{Q}_{k}\mathbf{H}^{\cH}_{k,k}+\mathbf{R}_{k,k}\right)^{-1}\right\}\label{P10ObjFunc}\\
\st \quad &\textrm{Tr}\{\mathbf{Q}_{k}\} \le p^{\textrm{max}}_{k}, \quad k \in \cK~\label{P10TxPower}\\
& \textrm{Tr}\{\mathbf{G}_{k}\mathbf{Q}_{k}\mathbf{G}^{\cH}_{k}\} \le t_{k},~
\forall~\Delta\mathbf{G}_{k} \in \mathcal{G}_{k}, k \in \cK~\label{P10RobustInterf}\\
& 0 \le t_k \le \iota_{k},\quad k \in \cK
\end{align}
\end{subequations}
which, as discussed in preceding sections, can be solved using the block coordinate ascent algorithm (or its proximal point version) in either a centralized or a distributed fashion.
After solving (P9) for a given set $\{\iota_{k}\}$, the per-CR interference budgets $\{\iota_{k}\}$ are updated by the following master problem:
\begin{subequations}
\begin{align}
\textrm{(P10)} \quad \mathop{\max}_{\{\iota_{k}\}}~&p(\{\iota_k\})\label{MasterObjFunc}\\
\st \quad &\{\iota_{k}\} \in \mathcal{I} \label{Mastercstt1}
\end{align}
\end{subequations}
with the simplex set $\mathcal{I}$ given by
\begin{align}
\label{eq:simplex}
\mathcal{I} := \left\{\{\iota_{k}\}|\iota_{k} \ge 0, \sum^{K}_{k=1}\iota_{k} \le \iota^{\textrm{max}}\right\} \, .
\end{align}

Overall, the primal decomposition method solves (P8) by iteratively solving (P9) and (P10).
Notice that the master problem (P10) dynamically divides the total interference budget $\iota^{\textrm{max}}$ among CR transmitters,
so as to find the best allocation of resources that maximizes the overall system performance.
Using the block coordinate ascent algorithm, the $k$-th transmit-covariance matrix $\mathbf{Q}_{k}$ is obtained by solving the following problem [cf. Algorithm~2]
\begin{subequations}
\begin{align}
\textrm{(P11)}\quad \tilde{p}_{k}(\iota_{k}) := \nonumber \\
\mathop{\max}\limits_{\mathbf{Q}_{k}\succeq \mathbf{0},t_k\ge 0}\,\,
&u_k\left(\mathbf{Q}_{k},\mathbf{Q}_{-k}\right)
+\Tr\left\{\mathbf{D}_{k}^{\cH}\mathbf{Q}_{k}\right\}~\label{P11ObjFunc}\\
\st \quad &\textrm{Tr}\{\mathbf{Q}_{k}\} \le p^{\textrm{max}}_{k}\label{P11TxPower}\\
& \textrm{Tr}\{\mathbf{G}_{k}\mathbf{Q}_{k}\mathbf{G}^{\cH}_{k}\} \le t_{k},
\forall~\Delta\mathbf{G}_{k} \in \mathcal{G}_{k}~\label{P11RobustInterf}\\
& t_{k} \le \iota_{k}~\label{P11LagMulti}
\end{align}
\end{subequations}
where the proximal point-based regularization term is added if Algorithm~2(P) is implemented.
Since (P11) is a convex problem, it can be seen that the subgradient of $\tilde{p}_{k}(\iota_{k})$
with respect to $\iota_{k}$ is the optimal Lagrange multiplier $\lambda_k$ corresponding to the constraint~\eqref{P11LagMulti}~\cite[Chap.~5]{Bertsekas}.
Thus, it becomes possible to utilize the subgradient projection method to solve the master problem.
Strictly speaking, due to the non-convexity of the original objective~\eqref{P9ObjFunc}, primal decomposition method leveraging the subgradient algorithm
is not an exact, but rather an approximate (and simple) approach to solve (P8).
However, because~\eqref{P11ObjFunc} is a \emph{tight} concave lower bound of~\eqref{P9ObjFunc} around the approximating feasible point,
$p(\{\iota_k\})$ is well-approximated by $\tilde{p}_{k}(\iota_{k})$ as $\{\mathbf{Q}_{k}^{(n)}\}$ approaches the optimal value $\{\mathbf{Q}_{k}^{\textrm{opt}}\}$. Hence, $\lambda_k$ also comes ``very close''
to the true subgradient of $p_{k}(\{\iota_{k}\})$ with respect to $\iota_k$.
Therefore, at iteration $\ell$ of the primal decomposition method, the subgradient projection updating the interference budgets $\bm \iota:= [\iota_1,\iota_2,\ldots,\iota_{K}]^{T}$ becomes
\begin{align}
\label{SubgradProj}
\bm{\iota}(\ell+1)= \textrm{Proj}_{\mathcal{I}} \left[\bm{\iota}(\ell)+ s(\ell)\bm{\lambda}(\ell)\right]
\end{align}
where $\bm{\lambda} := [\lambda_1,\lambda_2,\ldots,\lambda_{K}]^{T}$;
$s(\ell)$ is a positive step size; $\textrm{Proj}_{\mathcal{I}}[\cdot]$ denotes projection onto the convex feasible set $\mathcal{I}$.
Projection onto the simplex set in~\eqref{eq:simplex} is a computationally-affordable operation that can be efficiently implemented as in e.g.,~\cite{Palomar05}.

Once (P9) is solved distributedly, each CR that is coupled by the interference constraint has to transmit the local scalar Lagrange multiplier $\lambda_k(\ell)$ to a cluster-head CR node.
This node, in turn, will update $\{\iota_{k}(\ell+1)\}$ and will feed these quantities back to the CRs. The resulting on-line distributed scheme is tabulated as Algorithm~3.
Notice however that in order for the overall algorithm to adapt to possibly slowly varying channels, operation~\eqref{SubgradProj} can be computed at the end of each cycle
of the block coordinate ascent algorithm, rather than wait for its convergence.

%%%%%%%%%%%%%%%%%%%%%%%%%%%%%%%%%%%%%%%%%%%%%%%%%%%%
\begin{algorithm}[t]
\caption{Distributed on-line robust sum-MSE minimization with aggregate interference constraint}
\label{algo:iterSDP_primal}
\begin{algorithmic}[1]
\State Initialize $\mathbf{Q}_{k}^{(0)}(0)=\mathbf{0}$, and $\iota_k(0) = \iota^{\textrm{max}}/K$, $\forall~k \in \cK$
\Repeat \quad ($\ell = 1,2,\ldots$)
    \State [CRs]: Solve (P9) via Algorithm~2 [Algorithm~2(P)]
    \State [CRs]: Transmit $\{\lambda_k(\ell)\}$ to the cluster-head node
    \State [Cluster-head node]: Update $\{\iota_k(\ell+1)\}$ via~\eqref{SubgradProj}
    \State [CRs]: Receive $\{\iota_k(\ell+1)\}$ from the cluster-head node
\Until $u_k\left(\mathbf{Q}^{(n)}(\ell)\right) - u_k\left(\mathbf{Q}^{(n^{\prime})}(\ell-1)\right) < \upsilon,~\forall~k \in \cK$
\end{algorithmic}
\end{algorithm}
%%%%%%%%%%%%%%%%%%%%%%%%%%%%%%%%%%%%%%%%%%%%%%%%%%%%%%

% ----------------------------
\section{Simulations}
\label{sec:numericalresults}
% ----------------------------
In this section, numerical tests are performed to verify the
performance merits of the novel design.
The path loss obeys the model $d^{-\eta}$, with $d$ the
distance between nodes, and $\eta = 3.5$. A flat Rayleigh fading model
is employed. For simplicity, the distances of links $U_k^t \rightarrow U_k^r$
are all set to $d_{k,k} = 30$ m; for the interfering links $\{U_k^t \rightarrow U_j^r,j\neq k\}$
distances are uniformly distributed over the interval $30-100$~m.
As for the distances between CR transmitters and PU receivers, two different cases are considered:
(c1) the PU receivers are located at a distance from the CRs that is uniformly distributed over $70 - 100$~m; and,
(c2) the CR-to-PU distances are uniformly distributed over $30 - 100$~m.
Finally, the maximum transmit-power and the noise power are identical for all CRs.
For the proximal point-based algorithm, the penalty factors $\{\tau_k\}$
are selected equal to $0.1$.

\begin{figure}[t]
    \centering
    \subfigure[Case (c1)]
    {
        \includegraphics[width=0.45\textwidth]{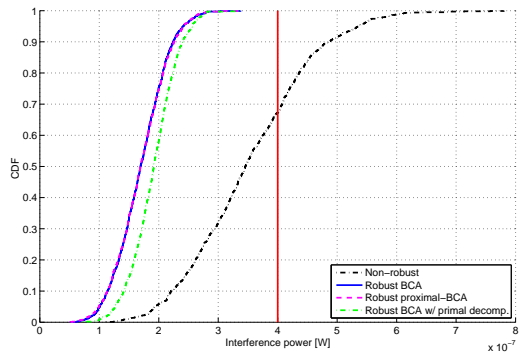}
        \label{F2_interfer_K4_222}
    }
    \subfigure[Case (c2)]
    {
        \includegraphics[width=0.45\textwidth]{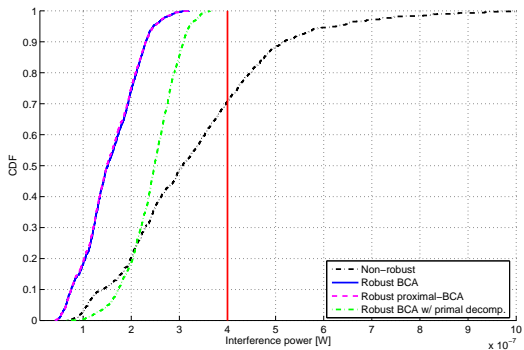}
        \label{F2_intermix_interfer_K4_222}
    }
    \caption{Interference cumulative distribution function (CDF).}
    \label{fig:Interf_CDF}
\end{figure}

To validate the effect of the robust interference constraint, the
cumulative distribution functions (CDF) of the interference power
at the PU are depicted in Fig.~\ref{fig:Interf_CDF}.
Four CR pairs and one PU receiver are considered,
all equipped with $2$ antennas.
The maximum transmit-powers and noise powers are set
so that the (maximum) signal-to-noise ratio (SNR) defined as
$\textrm{SNR} := p^{\textrm{max}}_{k}(d_{k,k}^{-\eta})/ \sigma^{2}_{k}$ equals $15$~dB. The total interference
threshold is set to $\iota^{\textrm{max}} = 4 \cdot 10^{-7}$~W and, for simplicity, it is equally split among the CR transmitters. The channel uncertainty is set to
$\epsilon^2_k = \rho \cdot \|\hat{\mathbf{G}}_{k}\|^2_F$~\cite{WangSP10}, with $\rho = 0.05$.
CDF curves are obtained using $2,000$ Monte Carlo runs. In each run, independent channel realizations are generated. The Matlab-based package \texttt{CVX}~\cite{CVX} along with \texttt{SeDuMi}~\cite{Sturm99} are used to solve the proposed robust beamforming problems.

The trajectories provided in Fig.~\ref{fig:Interf_CDF} refer to the block coordinate ascent (BCA) algorithm described in Section~\ref{Section:Robust}; the one with the proximal point-based regularization (proximal-BCA) explained in Section~\ref{sec:proximal}; and the non-robust solver of (P2), where the estimates  $\{\hat{\mathbf{G}}_{k}\}$ are used in place of the true channels $\{\mathbf{G}_{k}\}$. Furthermore, the green trajectory corresponds to (P8), where the subgradient projection~\eqref{SubgradProj} is implemented at the end of each BCA cycle, which includes $K$ updates of $\mathbf{Q}_k$ for $k=1,\ldots,K$. As expected, the proposed robust schemes enforce the interference constraint strictly in both scenarios (c1) and (c2). In fact, the interference never exceeds the
tolerable limit shown as the vertical red solid line in Fig.~\ref{fig:Interf_CDF}.
The CDFs corresponding to the proposed BCA and its proximal counterpart
nearly coincide. In fact, the two algorithms frequently converge
to identical stationary points in this particular simulation setup.
Notice that with the primal decomposition approach the beamforming strategy is less conservative.
On the contrary, the non-robust approach frequently violates the interference limit (more than $30\%$ of the time).
Finally, comparing Fig.~\ref{F2_interfer_K4_222} with Fig.~\ref{F2_intermix_interfer_K4_222},
one notices that the interference inflicted to the PU under (c1) and the one under (c2) are approximately of the same order. Since in the second case the CR-to-PU distances are smaller, the CR transmitters lower their transmit-powers to protect the PU robustly.

\begin{figure}[t]
    \centering
    \subfigure[Case (c1)]
    {
        \includegraphics[width=0.45\textwidth]{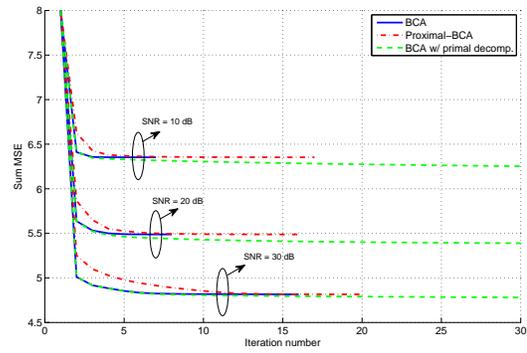}
        \label{F3_mse_K4_222}
    }
    \subfigure[Case (c2)]
    {
        \includegraphics[width=0.45\textwidth]{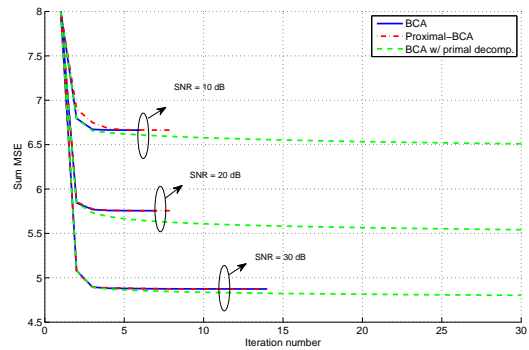}
        \label{F3_intermix_mse_K4_222}
    }
    \caption{Convergence of proposed algorithms, for SNR = 10, 20, and 30~dB.}
    \label{fig:MSEconverge}
\end{figure}

Convergence of the proposed algorithms with given channel realizations and over variable SNRs is illustrated in Fig.~\ref{fig:MSEconverge}.  It is clearly seen that the total MSEs decrease monotonically across fast-converging iterations, and speed is roughly identical in (c1) and (c2). As expected, the proximal point-based algorithm exhibits a slightly slower convergence rate.  Notice also that the primal decomposition method returns improved operational points, especially for medium and low SNR values. Furthermore, the gap between the sum-MSEs obtained with and without the primal decomposition scheme is more evident under (c2). Clearly, the sum-MSEs at convergence in (c2) are higher than the counterparts of (c1). This is because CRs are constrained to use a relatively lower transmit-power in order to enforce the robust interference constraints; this, in turn, leads to higher sum-MSEs and may reduce the quality of the CR-to-CR communications.

\begin{figure}[t]
\centering
\includegraphics[width=0.45\textwidth]{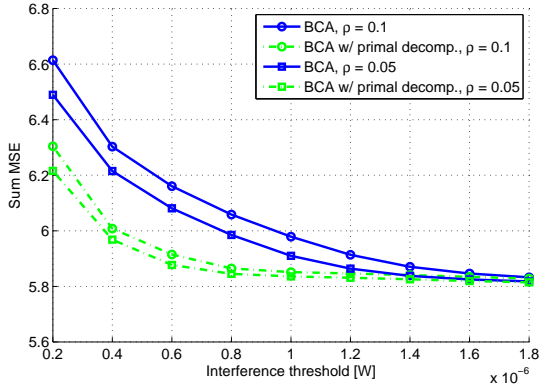}
\caption{Achieved sum-MSE as a function of $\iota^{\textrm{max}}$, for SNR = 10~dB.}
\label{F4_mse-interf_K4_222}
\end{figure}

\begin{figure}[t]
\centering
\includegraphics[width=0.45\textwidth]{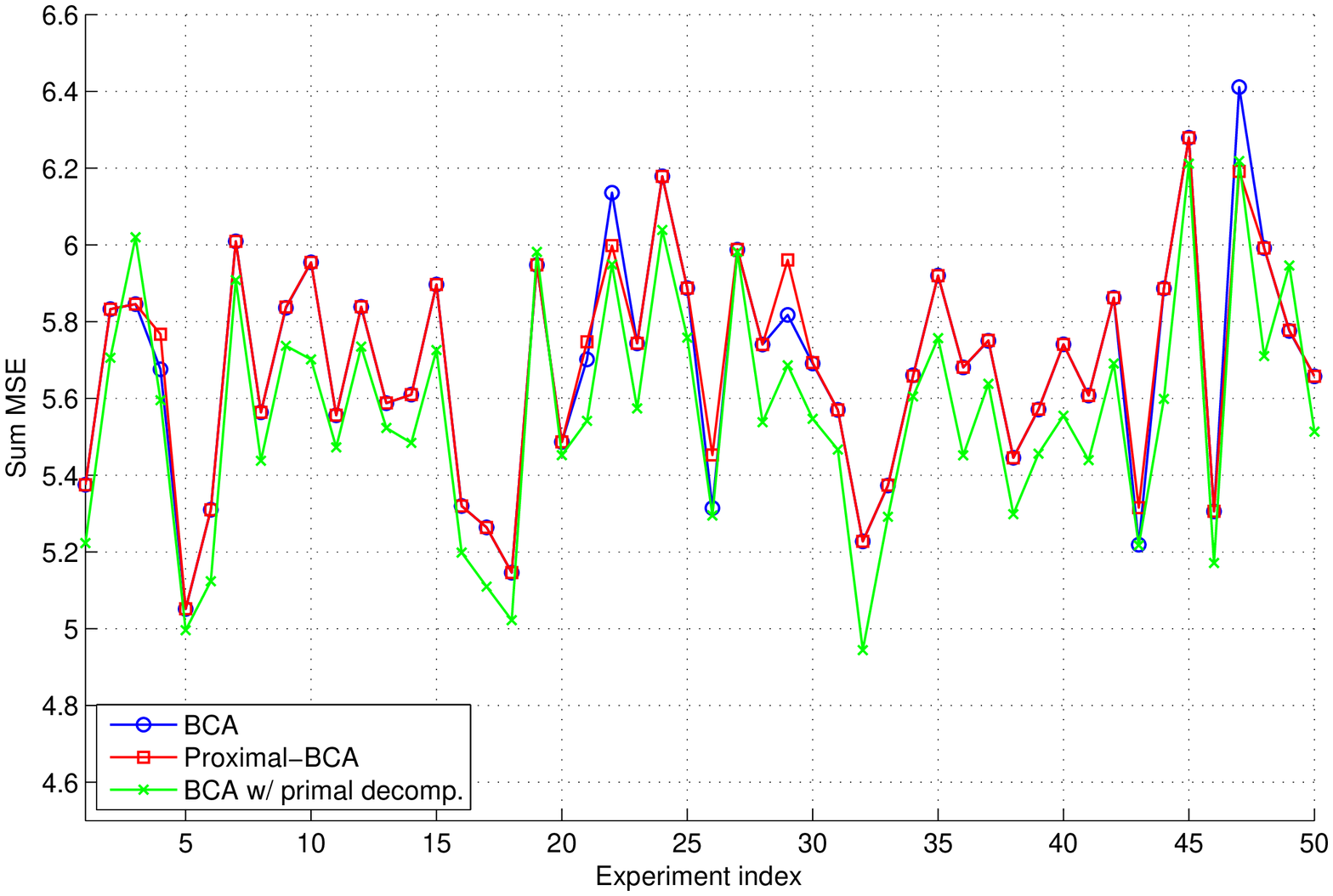}
\caption{Achieved sum-MSE for SNR = 15~dB.}
\label{F5_mse_50chs_K4_222}
\end{figure}

\begin{figure}[t]
\centering
\includegraphics[width=0.45\textwidth]{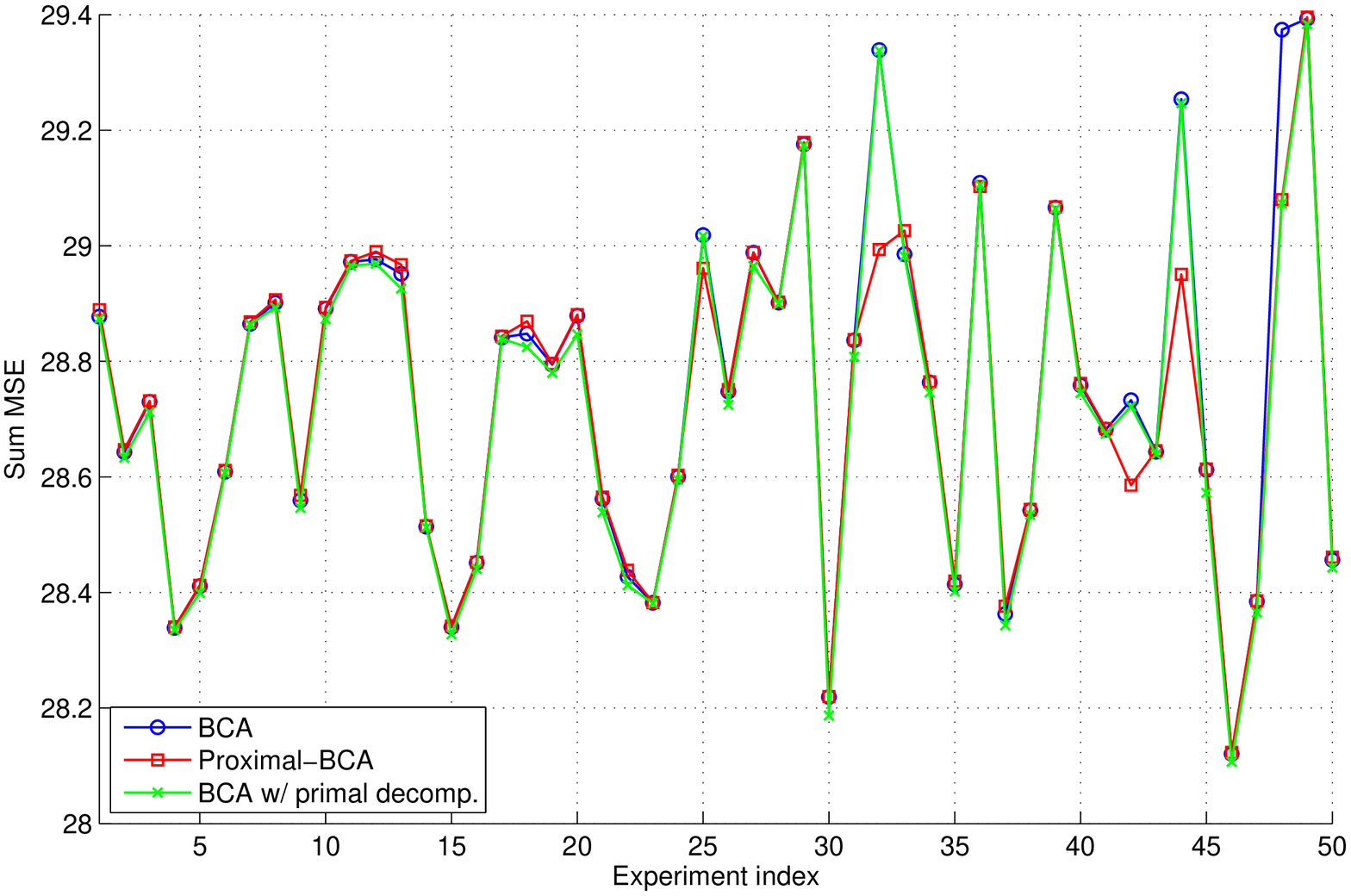}
\caption{Achieved sum-MSE for SNR = 15~dB.}
\label{F6_mse_50chs_K8_412}
\end{figure}

\begin{figure}[t]
\centering
\includegraphics[width=0.45\textwidth]{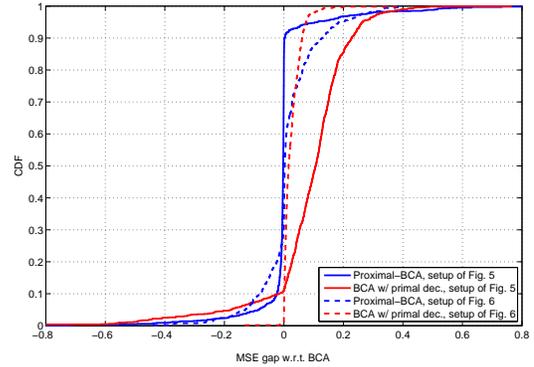}
\caption{CDF of sum-MSE gaps (relative to the BCA) using proximal-BCA (blue) and primal decomposition (red):
$\textrm{MSE}({\textrm{BCA}})- \textrm{MSE}(\textrm{proximal-BCA})$ and $\textrm{MSE}({\textrm{BCA}})- \textrm{MSE}(\textrm{BCA-primal decomp.})$.}
\label{F5_b_DiffCDF}
\end{figure}

In Fig.~\ref{F4_mse-interf_K4_222}, the achieved sum-MSE at convergence is reported as a function of the total interference threshold. Two sizes of the uncertainty region are considered with $\rho = 0.05$ and $\rho = 0.1$. Focusing on the first case, it can be seen that the two achieved sum-MSEs first monotonically decrease as the interference threshold increases, and subsequently they remain approximately constant. Specifically, for smaller $\iota^{\textrm{max}}$, the transmit-CRs are confined to relatively low transmit-powers in order to satisfy the interference constraint. On the other hand, for high values of $\iota^{\textrm{max}}$, the interference constraint is no longer a concern, and the attainable sum-MSEs are mainly due to CR self-interference. Notice also that for $\rho = 0.1$ the sum-MSEs are clearly higher, although they present a trend similar to the previous case. This is because the uncertainty region in~\eqref{P4Robust} becomes larger, which results in a higher sum-MSE.

In order to compare performance of the proposed algorithms, the total MSE obtained at convergence is depicted in Fig.~\ref{F5_mse_50chs_K4_222} for $50$ different experiments.
In each experiment, independent channel realizations are generated. The SNR is set to $15$ dB.
It is clearly seen that the objectives values of the two proposed methods often coincide. The differences presented in a few experiments are caused by convergence to two different stationary points. In this case, it is certainly convenient to employ the first algorithm, as it ensures faster convergence (see Fig.~\ref{F3_mse_K4_222}) without appreciable variations in the overall MSE. Notice that a smaller mean-square error can be obtained by resorting to the primal decomposition technique.

In Fig.~\ref{F6_mse_50chs_K8_412}, the simulation setup involves $8$ CR pairs and one PU receiver.
The CR transmitters have $4$ antennas, while the receiving CRs and the PU are equipped with $2$ antennas. The distances $d_{k,k}$ are set to $50$ m, while $\{d_{k,j}\}_{k \neq j}$ distances are uniformly distributed in the interval between $30$ and $250$~m. Finally, CR-to-PU distances are uniformly distributed between $100$ and $200$~m. Clearly, matrices $\{\mathbf{H}_{k,k}\}$ here do not have full column rank. It is observed that about $10\%$ of the times the proximal point based algorithm yields smaller values of the sum-MSE than Algorithm~1. This demonstrates that
Algorithm~1 may not converge to a stationary point, or, it returns an MSE that is likely to be worse than that of the proximal point-based scheme.

Fig.~\ref{F5_b_DiffCDF} depicts the CDFs of the difference between the sum-MSE obtained with BCA, along with the ones obtained with proximal-BCA and with the primal decomposition method. The simulation setups of Figs.~\ref{F5_mse_50chs_K4_222} and~\ref{F6_mse_50chs_K8_412} are considered. In the first case, it can be seen that for over $80\%$ of the trials the BCA and proximal-BCA methods yield exactly the same solution. Moreover, BCA with primal decomposition performs better than the BCA method about $90\%$ of the time. Specifically, the gain can be up to $0.765$, which corresponds to approximately $14\%$ of the average sum MSE of the BCA. In the second case, the proximal-BCA returns a smaller sum-MSE with higher frequency.

% ----------------------------
\section{Concluding summary}
\label{sec:conclusions}
% ----------------------------
Two beamforming schemes were introduced for underlay
MIMO CR systems in the presence of uncertain CR-to-PU propagation channels.
Robust interference constraints were derived by employing a norm-bounded
channel uncertainty model, which captures errors in the channel estimation phase,
or, random fading effects around the deterministic path loss. Accordingly, a robust
beamforming design approach was formulated to minimize the total MSE
in the information symbol reconstruction, while ensuring
protection of the primary system. In order to solve the formulated non-convex
optimization problem, a cyclic block coordinate ascent
algorithm was developed, and its convergence to a stationary
point was established when all CR-to-CR direct channel matrices have full column rank.
A second algorithm based on a proximal point regularization technique was also developed.
Although slower than the first, the proximal point-based scheme
was shown capable of converging to a stationary point even for rank-deficient channel matrices.
The two solutions offer complementary strengths as far as convergence rate, computational complexity,
and MSE optimality are concerned. They can both afford on-line distributed implementations.
Finally, a primal decomposition technique was employed to approximately solve the robust beamforming problem with coupled interference constraints.
The developed centralized and distributed algorithms are also suitable for non-CR MIMO ad-hoc networks as well as for conventional downlink or uplink multi-antenna cellular systems.

\appendix

\subsection{Derivation of the complex gradient matrix~\eqref{eq:gradD}}
\label{Appendix:gradD}
Using that $\frac{\partial \Tr\left\{\mathbf{X}^{\cH}\mathbf{A}\right\}}{\partial\mathbf{X}^{*}}=\mathbf{A}$~\cite{Magnus99},
and letting $[\mathbf{A}]_{mn}$ denote the $(m,n)$-th entry of matrix $\mathbf{A}$, it follows that
\begin{align}
\frac{\partial [\mathbf{B}_j^{\cH}]_{st}}{\partial\mathbf{Q}_{k}^{*}}
= \frac{\partial \Tr\left\{\mathbf{e}_{s}^{\cH}\mathbf{H}_{j,k}\mathbf{Q}_{k}^{\cH}\mathbf{H}_{j,k}^{\cH}\mathbf{e}_{t}\right\}}
{\partial\mathbf{Q}_{k}^{*}}
= \mathbf{H}_{j,k}^{\cH}\mathbf{e}_{t}\mathbf{e}_{s}^{\cH}\mathbf{H}_{j,k}
\end{align}
\begin{align}
\frac{\partial [\mathbf{B}_j^{\cH}]_{st}}{\partial [\mathbf{Q}_{k}^{*}]_{mn}}
= \mathbf{e}_{m}^{\cH} \mathbf{H}_{j,k}^{\cH}\mathbf{e}_{t}\mathbf{e}_{s}^{\cH}\mathbf{H}_{j,k} \mathbf{e}_{n}
= \mathbf{e}_{s}^{\cH}\mathbf{H}_{j,k} \mathbf{e}_{n}\mathbf{e}_{m}^{\cH} \mathbf{H}_{j,k}^{\cH}\mathbf{e}_{t}
\end{align}
which can be written in a compact form as
$\frac{\partial \mathbf{B}_j^{\cH}}{\partial [\mathbf{Q}_{k}^{*}]_{mn}}
= \mathbf{H}_{j,k}\mathbf{e}_{n}\mathbf{e}_{m}^{\cH}\mathbf{H}_{j,k}^{\cH}$.
Then, the identity
$\frac{\partial f}{\partial \mathbf{X}^{*}}
= -\mathbf{X}^{-1}\left(\frac{\partial f}{\partial (\mathbf{X}^{-1})^{*}}\right)\mathbf{X}^{-1}$~\cite{Magnus99}, which holds for any Hermitian positive definite matrix $\mathbf{X}$, is used to obtain
\begin{align}
\frac{\partial u_j}{\partial \mathbf{B}_{j}^{*}}
= -\mathbf{B}_{j}^{-1}\frac{\partial \Tr\left\{\mathbf{V}_{j}(\mathbf{B}_{j}^{-1})^{\cH}\right\}}
{\partial (\mathbf{B}_{j}^{-1})^{*}}\mathbf{B}_{j}^{-1}
= -\mathbf{B}_{j}^{-1}\mathbf{V}_{j}\mathbf{B}_{j}^{-1}.
\end{align}
Using now the chain rule, one arrives at
\begin{align}
\frac{\partial u_j}{\partial [\mathbf{Q}_{k}^{*}]_{mn}}
&= \Tr\left\{
\left(\frac{\partial u_j}{\partial\mathbf{B}_{j}^{\cH}}\right)^{T}
\frac{\partial \mathbf{B}_j^{\cH}}{\partial [\mathbf{Q}_{k}^{*}]_{mn}}
\right\} \nonumber \\
&=\Tr\left\{
-\mathbf{e}_{m}^{\cH}\mathbf{H}_{j,k}^{\cH} \mathbf{B}_{j}^{-1}
\mathbf{V}_{j}\mathbf{B}_{j}^{-1}\mathbf{H}_{j,k}\mathbf{e}_{n}
\right\}
\end{align}
which readily leads to the desired result
\begin{align}
\frac{\partial u_j}{\partial \mathbf{Q}_{k}^{*}}
= -\mathbf{H}^{\cH}_{j,k}\mathbf{B}_{j}^{-1}\mathbf{V}_{j}
\mathbf{B}_{j}^{-1}\mathbf{H}_{j,k}.
\end{align}

\subsection{Proof of Lemma~\ref{lemma:convexfunc}}\label{Appendix:convexfunc}
First, convexity of $\mathcal{Q}_{k}$ can be readily proved by the definition of a convex set~\cite[Ch.~2]{Boyd}.
Re-write the function $u_{j}(\mathbf{Q}_{k},\mathbf{Q}_{-k})$ as [cf.~\eqref{per-utility},~\eqref{eq:matV}]
\begin{align}
u_{j}(\mathbf{Q}_{k},\mathbf{Q}_{-k}) =
\Tr\left\{\mathbf{V}_{j}^{1/2}\mathbf{P}_{j}^{-1}(\mathbf{Q}_{k})\mathbf{V}_{j}^{1/2}\right\}
\end{align}
where
\begin{align*}
\mathbf{P}_{j}(\mathbf{Q}_{k}) = \mathbf{H}_{j,k}\mathbf{Q}_{k}\mathbf{H}_{j,k}^{\cH}
+ \sum_{i \neq k}\mathbf{H}_{j,i}\mathbf{Q}_{i}\mathbf{H}_{j,i}^{\cH} + \sigma_j^2\mathbf{I}_{N_j}
\end{align*}
is an affine map with respect to $\mathbf{Q}_{k}$.
Since $u_{j}$ is convex in $\mathbf{P}_{j}$~\cite[Theorem~2]{Lieb73}, and convexity is preserved under affine mappings and
nonnegative weighted-sums~\cite[Ch.~3]{Boyd}, it follows that $f_{k}(\mathbf{Q}_{k},\mathbf{Q}_{-k})$ is convex in $\mathbf{Q}_{k}$.

\subsection{Proof of Lemma~\ref{lemma:strictconcave}}\label{Appendix:strictconcave}
First, notice that the objective function~\eqref{P5ObjFunc} can be re-written as
\begin{align}
\widetilde{\cU}_{k}(\mathbf{Q}_{k}) = &N_k +\Tr\left\{\mathbf{D}_{k}^{\cH}\mathbf{Q}_{k}\right\}\nonumber \\
&-\Tr\left\{\mathbf{R}_{k,k} \left(\mathbf{H}_{k,k}\mathbf{Q}_{k}\mathbf{H}^{\cH}_{k,k}+\mathbf{R}_{k,k}\right)^{-1}\right\}.
\label{reformCost}
\end{align}
Then, it suffices to prove strict convexity in $\mathbf{Q}_{k}$ of the third term on the right hand side of~\eqref{reformCost}.
This is equivalent to showing that (subscripts are dropped for brevity)
\begin{align}
J(t) := \Tr\left\{\mathbf{R}\left(\mathbf{H}\mathbf{Q}\mathbf{H}^{\cH}+\mathbf{R}\right)^{-1}\right\}
\end{align}
is strictly convex in $t \in \{t|\mathbf{Q}:= \mathbf{X}+t\mathbf{Y} \in \cQ\}$ for any given
$\mathbf{X} \in \mathbb{H}^{n \times n}_{+}$ and nonzero $\mathbf{Y} \in \mathbb{H}^{n \times n}$.

To this end, consider the second-order derivative of ${J}(t)$, which is given by
\begin{align}
\ddot{J}(t) =
2\Tr\{\mathbf{C}\mathbf{R}\mathbf{C}\mathbf{L}\mathbf{C}\mathbf{L}\}\label{2ndDrvtvJ}
\end{align}
where
$\mathbf{C} := \left(\mathbf{R}+\mathbf{H}(\mathbf{X}+t\mathbf{Y})\mathbf{H}^{\cH}\right)^{-1}$
and $\mathbf{L} := \mathbf{H}\mathbf{Y}\mathbf{H}^{\cH}$. Note that matrix $\mathbf{C}\mathbf{R}\mathbf{C}$ is Hermitian positive definite, since
$\mathbf{C}$ and $\mathbf{R}$ are Hermitian positive definite too. With $\mathbf{H}$ full column rank, it readily follows that
$\mathbf{L} \neq \mathbf{0}$ for any $\mathbf{Y} \neq \mathbf{0}$. This ensures that the Hermitian positive semi-definite matrix $\mathbf{L}\mathbf{C}\mathbf{L}$
is not an all-zero matrix, i.e.,~$\mathbf{L}\mathbf{C}\mathbf{L}\neq \mathbf{0}$.

Let $\nu_{1}\ge\nu_{2}\ge \cdots \ge \nu_{N}>0$ and $\mu_{1}\ge\mu_{2}\ge \cdots \ge \mu_{N}\ge 0$
denote the eigenvalues of matrices $\mathbf{C}\mathbf{R}\mathbf{C}$ and $\mathbf{L}\mathbf{C}\mathbf{L}$, respectively.
Since matrix $\mathbf{L}\mathbf{C}\mathbf{L} \neq \mathbf{0}$, $\mu_{1}$ is strictly positive, and thus
\begin{subequations}
\begin{align}
\ddot{J}(t) &\ge 2\sum_{i=1}^{N}\nu_{i}\mu_{N-i+1}\label{VonNeumannIneq}\\
&\ge 2\nu_{N}\mu_{1} >0\label{eq:Jstrongcnvx}
\end{align}
\end{subequations}
where~\eqref{VonNeumannIneq} follows from von Neumann's trace inequality~\cite{Mirsky59}.
Finally, ~\eqref{eq:Jstrongcnvx} shows the strong convexity (and hence strict convexity)
of $J(t)$.

For completeness, we provide an alternative proof of the lemma. With some manipulations,
function $h(\mathbf{Q}):=\Tr\left\{\mathbf{R}\left(\mathbf{H}\mathbf{Q}\mathbf{H}^{\cH}+\mathbf{R}\right)^{-1}\right\}$ can be re-expressed as
\begin{align}
h(\mathbf{Q})&= g(\mathbf{R}^{-1/2}\mathbf{H}\mathbf{Q}\mathbf{H}^{\cH}\mathbf{R}^{-1/2})\nonumber \\
&=\Tr\left\{\left(\mathbf{I}+\mathbf{R}^{-1/2}\mathbf{H}\mathbf{Q}\mathbf{H}^{\cH}\mathbf{R}^{-1/2}\right)^{-1}\right\}\label{eq:LinearMap}
\end{align}
where $g(\mathbf{X}):=\Tr\left\{\left(\mathbf{I}+\mathbf{X}\right)^{-1}\right\}$. Let $\lambda_{1}(\mathbf{X}), \ldots, \lambda_{n}(\mathbf{X})$
denote again the eigenvalues of a matrix $\mathbf{X}$. Note that the spectral function
$g(\mathbf{X}) = s(\lambda(\mathbf{X})):=\sum_{i}\left(\frac{1}{1+\lambda_{i}(\mathbf{X})}\right)$
is strictly convex if and only if the corresponding symmetric function $s(\cdot)$ is strictly convex~\cite{Davis57}.
To this end, the strict convexity of $\frac{1}{1+x}$ for $x\ge 0$ implies the strict convexity of
$s(\cdot)$, and thus of $g(\mathbf{X})$.
Under the condition of full column rank of $\mathbf{H}$, we will show that \emph{strict} convexity is preserved under the linear mapping in~\eqref{eq:LinearMap}.
Specifically, define
\begin{align*}
\check{\mathbf{Q}}_{i} := \mathbf{R}^{-1/2}\mathbf{H}\mathbf{Q}_{i}\mathbf{H}^{\cH}\mathbf{R}^{-1/2},\, i = 1,2.
\end{align*}
Then, for any $\mathbf{Q}_{1}\neq\mathbf{Q}_{2}\in \cQ$ and $0<\lambda<1$, we have that
\begin{subequations}
\begin{align}
h(\lambda\mathbf{Q}_{1}+(1-\lambda)\mathbf{Q}_{2}) &=  g(\lambda \check{\mathbf{Q}}_{1}+(1-\lambda) \check{\mathbf{Q}}_{2})\\
&<\lambda g(\check{\mathbf{Q}}_{1})+(1-\lambda)g(\check{\mathbf{Q}}_{2})\label{eq:gStrictConvex}\\
&=\lambda h(\mathbf{Q}_{1}) + (1-\lambda)h(\mathbf{Q}_{2})
\end{align}
\end{subequations}
where~\eqref{eq:gStrictConvex} follows from the strict convexity of $g(\cdot)$, and the fact that
$\check{\mathbf{Q}}_{1}\neq \check{\mathbf{Q}}_{2}$ holds for any $\mathbf{Q}_{1}\neq \mathbf{Q}_{2}$, since $\mathbf{H}$ is full column rank.

%%%%%%%%%%%%%%%%%%%%%%%%%%%%%%%%%%%%%%%%%%%%%%
\bibliographystyle{IEEEtran}
\bibliography{biblio}

\end{document}